\documentclass[a4paper,fleqn,usenatbib]{mnras}
\usepackage{newtxtext,newtxmath}
\usepackage{graphicx}
\usepackage[T1]{fontenc}
\usepackage{ae,aecompl}
\usepackage{color}

\newcommand{\kms}{km~s$^{-1}$}
\newcommand\solmass{$\rm M_{\sun}$}
\newcommand\solum{$\rm L_{\sun}$}

\newcommand{\atlas}{{\sc Atlas$^\mathrm{3D}$}}
\newcommand{\e}[1]{$\times 10^{#1}$}
\newcommand{\cii}{{[C\,\sc ii]}}
\newcommand{\oi}{{[O\,\sc i]}}
\newcommand{\oiii}{{[O\,\sc iii]}}
\newcommand{\persqcm}{~cm$^{-2}$}
\newcommand{\jybkms}{Jy~beam$^{-1}~$km~s$^{-1}$}
\newcommand{\jykms}{Jy~km~s$^{-1}$}
\newcommand{\hi}{{H\,\sc i}}
\newcommand{\arcdeg}{\degr}
\newcommand{\nodata}{...}

\defcitealias{serra_vcirc}{S16}
\defcitealias{serra12}{S12}
\defcitealias{serra14}{S14}

\title[HI in slow rotator early-type galaxies]{The Diversity of Atomic Hydrogen in Slow Rotator Early-type Galaxies}

\author[L.~M.~Young et al.]{
Lisa M.\ Young,$^{1,2}$\thanks{E-mail: lisa.young@nmt.edu}
Paolo Serra,$^{3}$
Davor Krajnovi\'c,$^{4}$
and Pierre-Alain Duc$^{5}$
\\
$^{1}$Physics Department, New Mexico Tech, 801 Leroy Place, Socorro, NM 87801 USA\\
$^{2}$Adjunct Astronomer, National Radio Astronomy Observatory\\
$^{3}$INAF - Osservatorio Astronomico di Cagliari, Via della Scienza 5, I-09047 Selargius (CA), Italy\\
$^{4}$Leibniz-Institut f\"ur Astrophysik Potsdam (AIP), An der Sternwarte 16, D-14482 Potsdam, Germany\\
$^{5}$Observatoire Astronomique de Strasbourg, Universit\'e de Strasbourg, CNRS, UMR 7550, 11 rue de
l'Universit\'e, F-67000 Strasbourg, France
}

\date{Accepted XXX. Received YYY; in original form ZZZ}

\pubyear{2018}

\begin{document}
\label{firstpage}
\pagerange{\pageref{firstpage}--\pageref{lastpage}}
\maketitle

\begin{abstract}
We present interferometric observations of \hi\ in nine slow rotator early-type galaxies of the
\atlas\ sample.  With these data, we now have sensitive \hi\ searches in 34 of the 36 slow rotators.
The aggregate detection rate is 32\% $\pm$ 8\%, consistent with previous work;  however, we find two detections
with extremely high \hi\ masses, whose gas kinematics are substantially different from what was
previously known about \hi\ in slow rotators.
These two cases (NGC~1222 and NGC~4191) broaden the known diversity of \hi\ properties in slow rotators.
NGC~1222 is a merger remnant with prolate-like rotation and, if it is indeed prolate in shape, an equatorial gas
disc; NGC~4191 has two counterrotating stellar discs and an unusually large \hi\
disc.  We comment on the implications of this disc for the formation of $2\sigma$ galaxies.
In general, the \hi\ detection rate, the incidence of relaxed \hi\ discs, and the \hi/stellar mass
ratios of slow rotators are indistinguishable from those of fast rotators.
These broad similarities suggest
that the \hi\ we are detecting now is unrelated to the galaxies' formation processes and was often
acquired after their stars were mostly in place.
We also discuss the \hi\ nondetections; some of these galaxies that are undetected in
\hi\
or CO are detected in other tracers (e.g. FIR fine structure lines and dust).
The question of whether there is cold gas in
massive galaxies' scoured nuclear cores still needs work.
Finally, we discuss an unusual isolated \hi\ cloud with a surprisingly faint (undetected) optical
counterpart.
\end{abstract}


\section{Introduction}

Recent surveys for cold gas in early-type (elliptical and lenticular) galaxies have found surprising
numbers of detections.
For example, in a large compilation, \citet{a3d_cmd} found that at least 40\% of local early-type
galaxies have either atomic gas (\hi) or molecular gas (CO), or both, in masses of roughly $10^7$ \solmass\ or
greater.  These cold gas masses correspond to roughly $10^{-4}$ to $10^{-1}$ of their total stellar
masses.
The sample is from the \atlas\ project \citep{cappellari_a3d1}, which is a volume-limited
survey of early-type galaxies closer than about 40 Mpc and with stellar masses greater than about
$10^{9.9}$ \solmass; crucially, the sample is defined with a $K_s$ magnitude 
cut, which means that the sample is not biased towards younger stellar populations.
And these relatively high detection rates open up the possibility of using the cold gas as another key to the evolutionary
histories of many early-type galaxies.

To make use of this key, it is necessary to know where the cold gas came from.
Evidently, sometimes cold molecular gas has cooled and condensed out of a hot X-ray atmosphere
\citep[e.g.][]{lim_1275,temi2017}.
In one recent interesting case, \citet{russell2017} mapped CO emission in and around the central galaxy
of the Abell 1795 cluster.  They suggest that
relatively low-entropy hot gas was entrained and lifted by rising radio bubbles, and where the
hot gas was compressed around the edges of the bubbles, it became thermally unstable and cooled to
molecular temperatures.
But most early-type galaxies are not the central cluster-dominant or group-dominant members, so
additional mechanisms are also needed to explain the majority of the \hi\- and CO-detected early-types.
Sometimes kinematic misalignments between the gas and the stars show that the cold gas was probably
acquired relatively late in the galaxy's history, after the bulk of the stars formed.
In those cases the gas could have been accreted from a filament and/or from a cannibalized satellite
galaxy.  In other cases the cold gas could be recycled (cooled) stellar mass loss material
\citep[e.g.][]{davis11,serra14,negri2015}.

Meanwhile, it is also evident that early-type galaxies come in several different flavors that are
believed to have different formation histories \citep{cappellari_araa}.
The most massive early-types tend to be slow rotators, spherical or weakly triaxial in shape, and 
probably formed through a series of gas-poor mergers at the centers of group and cluster halos.
Slow rotators at lower masses can also form through major binary mergers that have a particular
geometry with a small net angular momentum \citep[see also][and references therein]{bois,lagos2017}. 
Fast rotators were disc galaxies at high or even moderate redshift, but they have lost a good deal (not
necessarily all) of their cold gas and have suffered various disc heating and bulge growth processes as
their stellar populations have reddened.  Fast rotators can also form through major binary mergers with
a particular geometry that gives a high net angular momentum.

In this context, the broad motivation of this project is to 
make connections between the histories of early-type galaxies as encoded in their stars
and in their cold gas.
Studies of the quantities and kinematics of cold gas in early-type galaxies can help clarify the nature
of the environmental and internal quenching processes that move them towards the red sequence.
They can help quantify
the potential for rejuvenation episodes, such as the accretion of cold gas and the
growth of a young stellar disc in a spheroid-dominated galaxy. 
And \hi\ in particular often exists at large radii, where it can provide evidence of past interactions
for long times \citep[e.g.][]{freeke_misalign}.

Here we
present archival and new data on \hi\ emission in nine slow rotator early-type galaxies; with these
additions,
virtually all of the slow rotators in the volume-limited \atlas\ survey now have sensitive \hi\ data.
In Sections \ref{sec:sample} and \ref{sec:data} we describe the selection of targets, the observations,
and the data reduction.
In Section \ref{sec:results_general} we provide updated statistics on \hi\ in slow rotators, compare the
properties of their \hi\ to fast rotators, and consider some implications for the origin of the gas.
In Section \ref{sec:nondets} we provide some additional notes on the galaxies not detected in \hi.
In Sections \ref{sec:1222} and \ref{sec:4191} two galaxies are discussed in more detail, presenting
their \hi\ kinematics, its relation to the stellar kinematics and dynamics, and some ideas about whether
the galaxies were formed in recent wet mergers or accreted their large gas contents from other sources.
Section \ref{sec:summary} summarizes this material on \hi\ in slow rotators, and the Appendix presents a basic
compilation of \hi\ detections in other spirals and dwarf galaxies that happened to be in the observed
fields.  One particularly interesting \hi\ detection does not have a known optical companion.

\section{SAMPLE}\label{sec:sample}

The specific motivation for this
project was to obtain more complete information on the \hi\ content of the slow rotators in the \atlas\
sample.  Slow rotators are intrinsically rare, and relatively few of them have been observed or
detected in \hi\ emission.  Specifically,
the \atlas\ survey contained a total of 224 fast rotators \citep{emsellem_a3d}, of which 111 were observed
in \hi\ emission with the WSRT by \citet[hereafter S12]{serra12} and 41 were detected.
The survey also contained 36 slow rotators, of which 21 were observed with WSRT, 
one was observed with Arecibo, and 9 were detected.
The remaining 14 slow rotators were not discussed in \citetalias{serra12} because they
are at sufficiently low Declination to make WSRT observations difficult or because
they are members of the Virgo Cluster.  

Analyzing these data, \citet[hereafter S14]{serra14} found relatively few differences between
the \hi\ properties of fast and slow rotators; their \hi\ detection rates, masses, column densities, and
morphologies are similar.  These similarities are curious and informative because the two types of
galaxies probably have very different assembly histories \citep[as mentioned above;][]{cappellari_araa}.  
However, it is notable that the slow rotator class is internally
diverse \citep{davor} and nine \hi\ detections is not a large number.  
Thus it would be beneficial to
assemble the \hi\ information for the remaining slow rotators.

These remaining slow rotators are easily accessible to the NSF's Karl G.\ Jansky Very Large Array (VLA), and in some instances they have \hi\
observations from Arecibo.
One of them, NGC~4476, has a previously published \hi\ nondetection from the VLA
\citep{lucero4476}.
Four of them (NGC~1222, NGC~1289, NGC~5831, and NGC~5846) have archival VLA data which has 
not been adequately published elsewhere, and it is presented here.  Five of them (NGC~4191,
NGC~4636, NGC~4690, NGC~5576, and NGC~5813) are newly observed with the VLA and also presented here.
In two cases we have obtained \hi\ upper limits from the Arecibo observations of the ALFALFA
survey \citep{giovanelli05}.  The remaining two, NGC~4486 and NGC~4261, will always be
difficult for \hi\ emission searches due to their extremely bright radio continuum, though
\citet{dwarakanath} discuss \hi\ upper limits in NGC~4486 from a much higher angular resolution
absorption experiment.

While the original targets were all slow rotators in the \atlas\ sample, the clustering nature of
galaxies means that some of the \atlas\ fast rotators and numerous late-type and dwarf galaxies were
also present in the observed VLA fields.  Table~\ref{tab:obs} lists the \atlas\ members that were
newly observed at the VLA for this project or retrieved from the archive, and additional information on the late-type and dwarf galaxies is given in Section
\ref{sec:detsnlims} and the Appendix.

\section{DATA AND IMAGING}\label{sec:data}

NGC~4191, NGC~4636, NGC~5576, and NGC~5813 were observed with the VLA
in October 2015, and NGC~4690 was observed in March 2017.  These data were obtained in the D configuration, giving baselines from 0.2 to 5
k$\lambda$ and resolutions $\sim$ 55\arcsec; more detailed observing parameters are given in Table~\ref{tab:obs}.
The array configuration and time on source were chosen to give beam sizes and sensitivities
close to those obtained at WSRT for the other \atlas\ galaxies.
The flux, bandpass, and time-dependent gain calibrations were made following standard
procedures, using 3C286 once per observing session as the primary flux
and bandpass calibrator, and secondary gain calibrators were
observed every 20 minutes.  
Calibration was performed using the eVLA scripted pipeline versions 1.3.8 and 1.3.9 in their
respective versions of the CASA software package 4.6.0 and 4.7.0.
The bootstrapped flux densities of the secondary gain calibrators are in good agreement
with previous measurements in the VLA calibrator database and the NRAO VLA Sky Survey \citep{nvss}.
An initial round of imaging revealed any bright line emission in the data, and the velocities
of that emission combined with the known target velocities were used to select frequency 
ranges for estimating continuum levels.
Continuum emission was subtracted using order 0 or 1 fits to the individual visibilities.
The final data cubes were produced with a Briggs robustness parameter of 0.5 and cleaned (where
necessary) to a residual level of 1 to 1.3 times the rms noise level.  

NGC~1222, NGC~1289, NGC~5831, and NGC~5846 were observed in several older VLA projects, as described in 
Table~\ref{tab:obs}.  Observing strategies, data reduction, and imaging are similar to those used for the
newer data, except as noted here.
The NGC~5846 group, which also includes NGC~5845, NGC~5838, and NGC~5831 from the \atlas\ sample, plus
many late-type galaxies, was observed in a square 36-point mosaic spaced at 15$\arcmin$ (the
half-power radius of the primary beam).  The final image has a gain above 0.90 in a square region
76\arcmin\ on a side, centered on 15h05m47 +01d34m30 (J2000).
We note that NGC~5839, also a member of the \atlas\ sample, is nominally in this field but it falls
outside the velocity range covered.
To the best of our knowledge, the archival data on NGC~1289 and the NGC~5846 group have not been
previously published.

In contrast to all the other data used here, NGC~1222 was observed in the C configuration, 
giving baselines from 0.3 to 16 k$\lambda$ and a resolution $\sim$ 20\arcsec. In addition, only 30
minutes were obtained on source; these data are thus higher resolution and lower sensitivity than
the other data discussed in this paper.  But as the \hi\ emission is strongly detected, that is
appropriate for this target. 
\citet{thomas04,thomas04b} present the total \hi\ flux and \hi\ column density from these data, but they 
do not
discuss the gas kinematics.   We have re-reduced the data with slightly higher resolution because we are
now able to make comparisons of the HI, CO, ionized gas and stellar kinematics, and that information
is not published elsewhere.

The sensitivity, final imaged velocity resolution and velocity coverage of these \hi\ observations are indicated 
in Table~\ref{tab:images}.
Following the discussion in \citetalias{serra12}, which uses data of similar sensitivity and
resolution, we indicate the \hi\ column density sensitivity as a 5$\sigma$ signal in one 
channel of width 16 \kms\ or as close to that as the data will allow.  \citetalias{serra12} also make a careful analysis of the typical angular sizes
and velocity widths of detected \hi\ features in the \atlas\ survey.  Based on this
analysis, they adopt M(\hi) upper limits for the \hi\ nondetections
by calculating the statistical uncertainty in a sum over a data volume of 50 \kms\ and six synthesized beam areas 
(typically 1.2\e{4} square arcseconds or 110 kpc$^2$ to 450 kpc$^2$ at the distances of this sample).
The mass limit corresponds to three times this formal uncertainty.
That procedure is also adopted here.
In Section \ref{stats} we also bring in some upper limits from the Arecibo telescope, and for those
data we adopt a mass limit defined by three times the uncertainty in a sum over 50 \kms\ and one Arecibo beam (3.7\e{4} square arcseconds).

\begin{table*}
{\centering
\caption{VLA Observations of \hi\ in \atlas\ galaxies.}
\label{tab:obs}
\begin{tabular}{lllcccc}
\hline
Target & Project & Obs. date & Config. & time on source & Bandwidth & Vel. resolution \\
       &         &           &         &  (hr)          & (MHz)  & (kHz) \\
\hline
NGC 1222 & AT0259 & Jul 2001 & C & 0.5 & 6.25 & 97.6 \\
NGC 1289 & AR0251 & Sep 1992 & D & 3  & 6.25 & 195.3 \\
NGC 4191 & 15B-258 & Oct 2015 & D & 1.5 & 16 & 15.6 \\
NGC 4636 & 14B-396 & Oct 2015 & D & 1.5 & 8 & 1.95 \\
NGC 4690 & 15B-258 & Mar 2017 & D & 1.5 & 16 & 15.6 \\
NGC 5576, 5574 & 15B-258 & Oct 2015 & D & 1.5 & 16 & 15.6 \\
NGC 5813 & 15B-258 & Oct 2015 & D & 1.5 & 16 & 15.6 \\
NGC 5846, 5845, 5838, 5831 & AZ0118 & Mar 1999 & D & 36 & 5.1 & 97.6 \\
\hline
\end{tabular}}
\end{table*}

\begin{table*}
{\centering
\caption{\hi\ image properties for Atlas3D galaxies.}
\label{tab:images}
\begin{tabular}{lllcccccccr}
\hline
Type & Target & Dist. & Vel.\ range & beam & beam & $\Delta v$ & rms & 
N(\hi) lim & S(\hi) & M(\hi) \\ 
     &        & (Mpc) &\ (\kms)     & (\arcsec) & (kpc) & (\kms) & (mJy/bm) &
($10^{19}$\persqcm) & (\jykms) & (\solmass) \\
(1) & (2) & (3) & (4) & (5) & (6) & (7) & (8) & (9) & (10) & (11)\\
\hline
 & NGC~1222 & 33.3 & 1865 -- 3060 & 20 $\times$ 15 & 3.2 $\times$ 2.4 & 21.0 & 1.15 & 43 & 11.2$\pm$0.7 & (2.9$\pm$0.2)\e{9} \\  
 & NGC~1289 & 38.4 & 2206 -- 3466 & 77 $\times$ 50 & 14 $\times$ 9.3 & 42.0 & 0.40 & 2.4 & $<$ 0.13 & $<$ 4.7\e{7} \\ 
 & NGC~4191  & 39.2 & 1132 -- 4172 &  49 $\times$ 45 & 9.3 $\times$ 8.6 & 16.0 & 0.70 & 2.7 & 10.1$\pm$0.3 & (3.7$\pm$0.1)\e{9} \\
 & NGC~4636  & 14.3 & 143  -- 1663 &  51 $\times$ 45 & 3.5 $\times$ 3.1 & 16.0 & 0.70 & 2.7 & $<$ 0.15 & $<$ 7.0\e{6} \\ 
Slow Rotators & NGC~4690  & 40.2 & 1216 -- 4264 & 70 $\times$ 50 & 14 $\times$ 9.7 & 16.0 & 0.80 & 2.0 & $<$ 0.16 & $<$ 6.3\e{7} \\
 & NGC~5576  & 24.8 & $-4$ -- 3020 & 54 $\times$ 46 & 6.5 $\times$ 5.5 & 16.0 & 0.65 & 2.3 & $<$ 0.14 & $<$ 2.0\e{7} \\
 & NGC~5813  & 31.3 & 421  -- 3453 &  51 $\times$ 46 & 7.7 $\times$ 7.0 & 16.0 & 0.67 & 2.5 & $<$ 0.14 & $<$ 3.2\e{7} \\ 
 & NGC~5831 & 26.4 & 1324 -- 2367 & 57 $\times$ 50 & 7.3 $\times$ 6.4 & 22.0 & 0.48 & 2.0 & $<$ 0.12 & $<$ 1.9\e{7} \\
 & NGC~5846 & 24.2 & 1324 -- 2367 & 57 $\times$ 50 & 6.7 $\times$ 5.8 & 22.0 & 0.48 & 2.0 & $<$ 0.12 & $<$ 1.6\e{7} \\
\hline
 & NGC~5574  & 23.2 & $-4$ -- 3020 & 54 $\times$ 46 & 6.1 $\times$ 5.2 & 16.0 & 0.65 & 2.3 & $<$ 0.14 & $<$ 1.7\e{7} \\
Fast Rotators & NGC~5838 & 21.8 & 1324 -- 2367 & 57 $\times$ 50 & 6.0 $\times$ 5.3 & 22.0 & 0.48 & 2.0 & $<$ 0.12 & $<$ 1.3\e{7} \\
 & NGC~5845 & 25.2 & 1324 -- 2367 & 57 $\times$ 50 & 7.0 $\times$ 6.1 & 22.0 & 0.48 & 2.0 & $<$ 0.12 & $<$ 1.8\e{7} \\ 
\hline
\end{tabular}
}

Distances are taken from \citet{cappellari_a3d1}.  The velocity
range in column 4 indicates the usable range covered by the data.  Column 7 gives the channel widths
in the image cubes used for analysis; for the newer data, these are significantly wider than the
intrinsic resolution of the data.  The column density limit represents 5$\sigma$ in one channel and
the integrated flux density limit represents three times the statistical uncertainty in a sum over 6
beams and 50 \kms. 
\end{table*}

\section{RESULTS: \hi\ in slow rotators}\label{sec:results_general}

\subsection{Summary of \hi\ detections and limits}\label{sec:detsnlims}

\begin{figure}
\includegraphics[scale=0.6, clip]{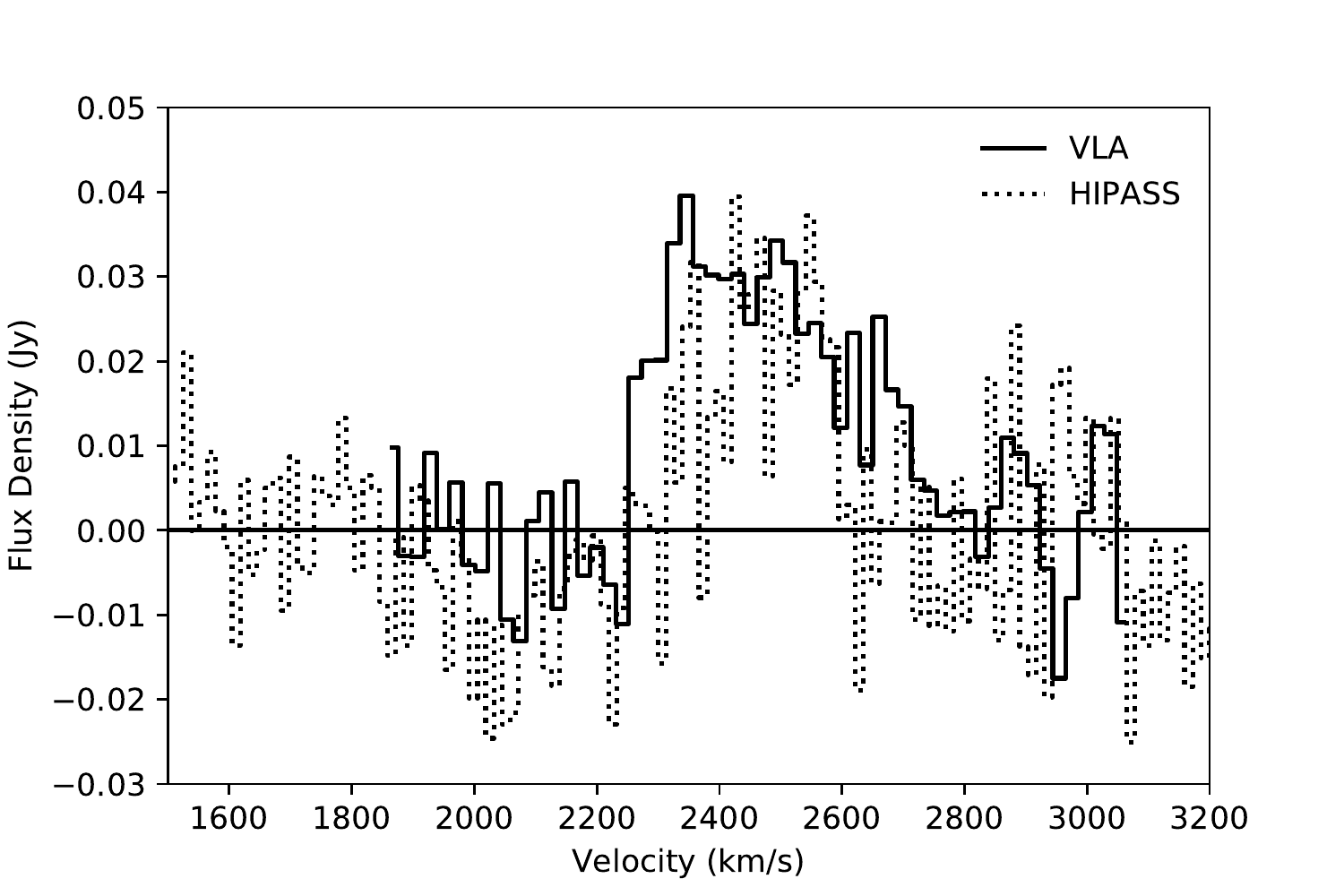}
\caption{Integrated \hi\ spectrum of NGC~1222, constructed by integrating each channel within a spatial
region defined in the column density (moment 0) image.  It is compared to the HIPASS specrum of
\citet{doyle05}, and the two spectra are consistent with each other given the
relatively low signal-to-noise ratio and baseline ripples in the HIPASS data.}
\label{fig:1222spec}
\end{figure}

\hi\ emission is detected in NGC~1222 at high signal-to-noise ratio, and more detailed descriptions of
the distribution and kinematics are provided in section \ref{sec:1222} below.
We measure the total line flux from an integrated spectrum (Figure~\ref{fig:1222spec}) to be 11.2
$\pm$ 0.7 $\pm 1$ \jykms, where the 0.7 \jykms\ represents the statistical uncertainty due to thermal
noise in the data and the 1 \jykms\ is a rough estimate of the uncertainty associated with the spatial
region of integration.  It is notable that 
the flux recovered in these data is 60\%
larger than the 6.9 $\pm$ 0.8 \jykms\ quoted by \citet{thomas04} using the same data.  
The discrepancy may be due to the choice of data volume to integrate,
as the recovered flux density of the gain calibrator B0320+053 is
consistent with the NRAO calibrator manual and the 1.4 GHz continuum flux density that we measure for NGC~1222
(61 $\pm$ 2 mJy) is consistent with the NVSS value of 62 $\pm$ 2 mJy \citep{nvss}.
For comparison, HIPASS and Effelsberg spectra show \hi\ fluxes of 9.4 \jykms\ and 9.5 $\pm$ 1.0 \jykms, respectively 
\citep{doyle05,hr89}, but both of the single dish spectra suffer from baseline ripples that are
probably related to the strong continuum emission.

\begin{figure}
\includegraphics[scale=0.6, clip]{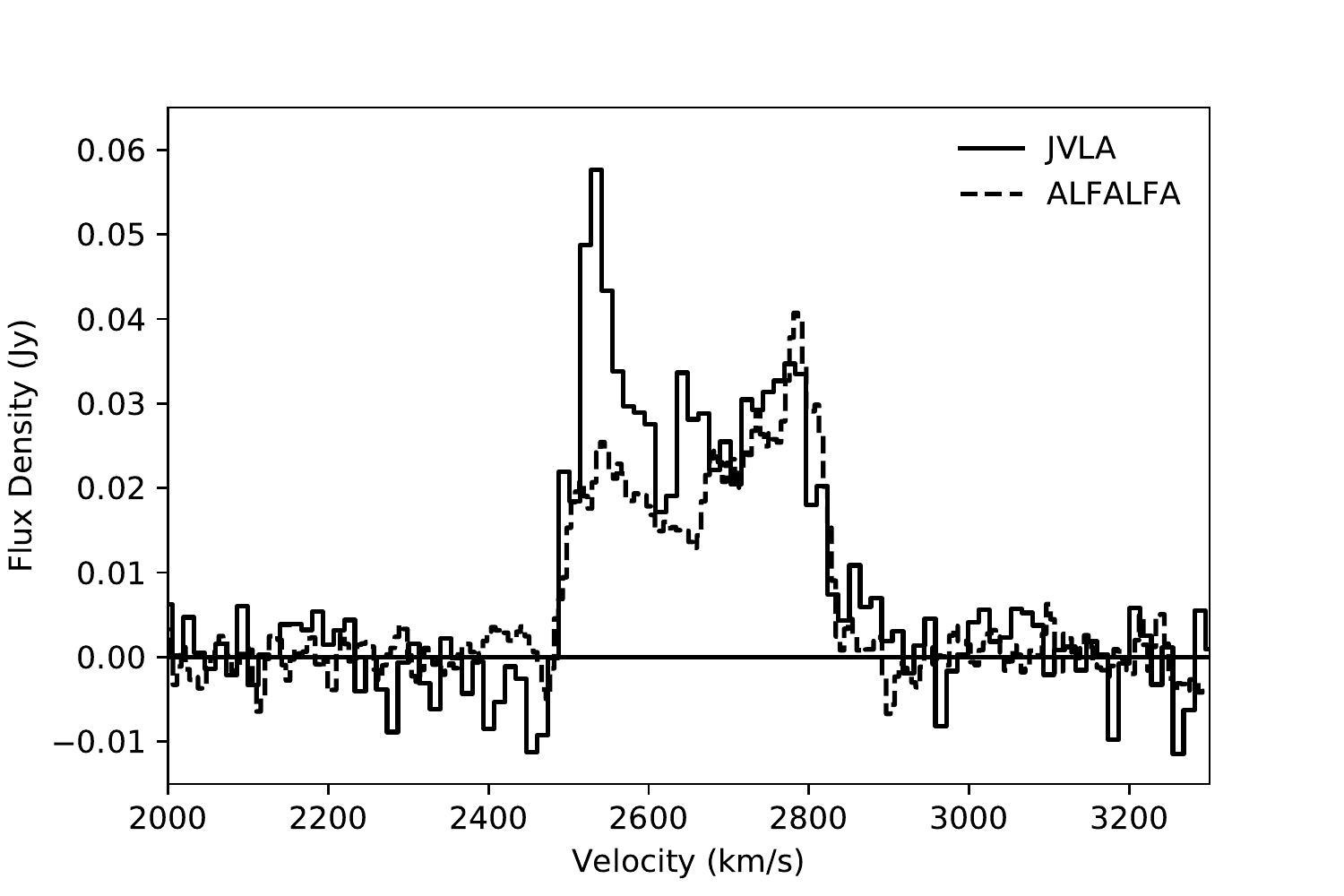}
\caption{Integrated \hi\ spectrum of NGC~4191, constructed by integrating each channel within a spatial
region defined in the column density (moment 0) image.  It is compared to the Arecibo specrum of
\citet{alfa40}, which misses most of the blue peak because the disc extends beyond the Arecibo
beam on the North side of the galaxy.}
\label{fig:4191spec}
\end{figure}

\hi\ emission is also strongly detected in NGC~4191; the integrated flux density recovered here, 10.1
$\pm$ 0.3 \jykms, is also significantly larger than the value of 7.52 \jykms\ reported by 
\citet{courtois&tully} based on Arecibo data \citep{alfa40}.  The two spectra are compared in Figure
\ref{fig:4191spec}.  The difference in flux is undoubtedly due to the
fact that the extent of the \hi\ emission is larger than the Arecibo beam, and further details on the
distribution and kinematics are in section \ref{sec:4191}.

No emission is found in the other \atlas\ targets, giving the mass limits in Table~\ref{tab:images}.
To the best of our knowledge, these are the first published \hi\ observations of 
NGC~1289, NGC~4690, NGC~5813, NGC~5831, and NGC~5845.
There are previous single-dish nondetections of 
NGC~4636, NGC~5574, NGC~5576, NGC~5838, and NGC~5846 \citep{duprie96,hr89}, though our new limits are
considerably more sensitive.  A few additional comments about these \hi\ non-detections are provided in
Section \ref{sec:nondets}.

In the field of NGC~5576 we do, however, detect significant emission from the spiral galaxy NGC~5577,
and we find another \hi\ detection whose optical counterpart is faint and not currently identified.
In the field of NGC~4690 we find emission from UGC~07945.
In the field of NGC~5813 we find emission from NGC~5806, NGC~5811, and UGC~09661, and in the large
mosaic of the NGC~5846 group there are \hi\ detections from six galaxies.
Additional information on those detections is in the Appendix, and
good upper limits could also be obtained for a large number of other objects in the mosaic
field.

\subsection{Updated statistics on \hi\ in slow rotators}\label{stats}

Table~\ref{tab:SRstats} presents the updated compilation of \hi\ masses or upper limits in the 
\atlas\ slow rotators.
Interferometric observations from WSRT or the VLA are now available for 31 of the 36 slow
rotators.  Three others have \hi\ upper limits from Arecibo.  NGC~4261 also has Arecibo
observations from the ALFALFA survey, but its bright continuum emission makes the noise
level at least a factor of 10 higher than for the other galaxies and we do not count it as
useful data.  The continuum issue also plagues \hi\ observations of NGC~4486, as mentioned in Section
\ref{sec:sample} above.

We thus have sensitive, reasonably high resolution \hi\ observations of 34 slow rotators and HI
detections in 11 of them (0.32 $\pm$ 0.08, for a 68\% confidence interval).
Outside of the Virgo Cluster, we have observations of 26 slow rotators and 10 detections
(0.38 $\pm$ 0.10).\footnote{The Virgo Cluster slow rotator detected in \hi\ is NGC~4406.}  
The new detection rates in slow rotators
have dropped by approximately 1 standard deviation compared to those reported in \citetalias{serra14},
because the newly compiled observations are preferentially nondetections.
Indeed, it seems probable that one could expect older but previously unpublished data to be
largely nondetections because of a publication bias.  

Completing the sample (or nearly so) gives much improved constraints on the incidence of \hi\ in slow
rotators, as the previous data left open the 
possibility that the \hi\ detection rate of slow rotators could be much (up to a factor of 2) higher than
that of the fast rotators.
Instead, the new results imply that the \hi\ detection rates of slow rotators are virtually
indistinguishable from those of fast rotators.
For specific examples, the overall \hi\ detection rate in the fast rotators in \citetalias{serra14} was
0.31 $\pm$ 0.05, comparable to the updated rate in slow rotators of 0.32 $\pm$ 0.08.
Considering only the galaxies outside of the Virgo Cluster, \citetalias{serra14} reported an \hi\ detection rate of 0.37
$\pm$ 0.06 for fast rotators and that value is in good agreement with the rate of 0.38 $\pm$ 0.10 for
slow rotators.

As an aside, we note that \citet{cappellari_araa} proposes a slightly modified definition of the
slow rotator class from the one we have adopted here.  Four of the 260 \atlas\ galaxies have different
classifications in the two systems.  In the modified definition, three of the slow rotators in Table
\ref{tab:SRstats} would become fast rotators.  These are NGC~3796, NGC~4550, and NGC~4690.  One new
slow rotator would be added to Table~\ref{tab:SRstats}; it is NGC~4733.  None of these are detected in
\hi\ emission and the \hi\ detection statistics would therefore not change significantly. 

\subsection{Additional comparisons of \hi\ properties in fast and slow rotators}

\begin{figure}
\includegraphics[scale=0.57, trim=1cm 1cm 1cm 1cm]{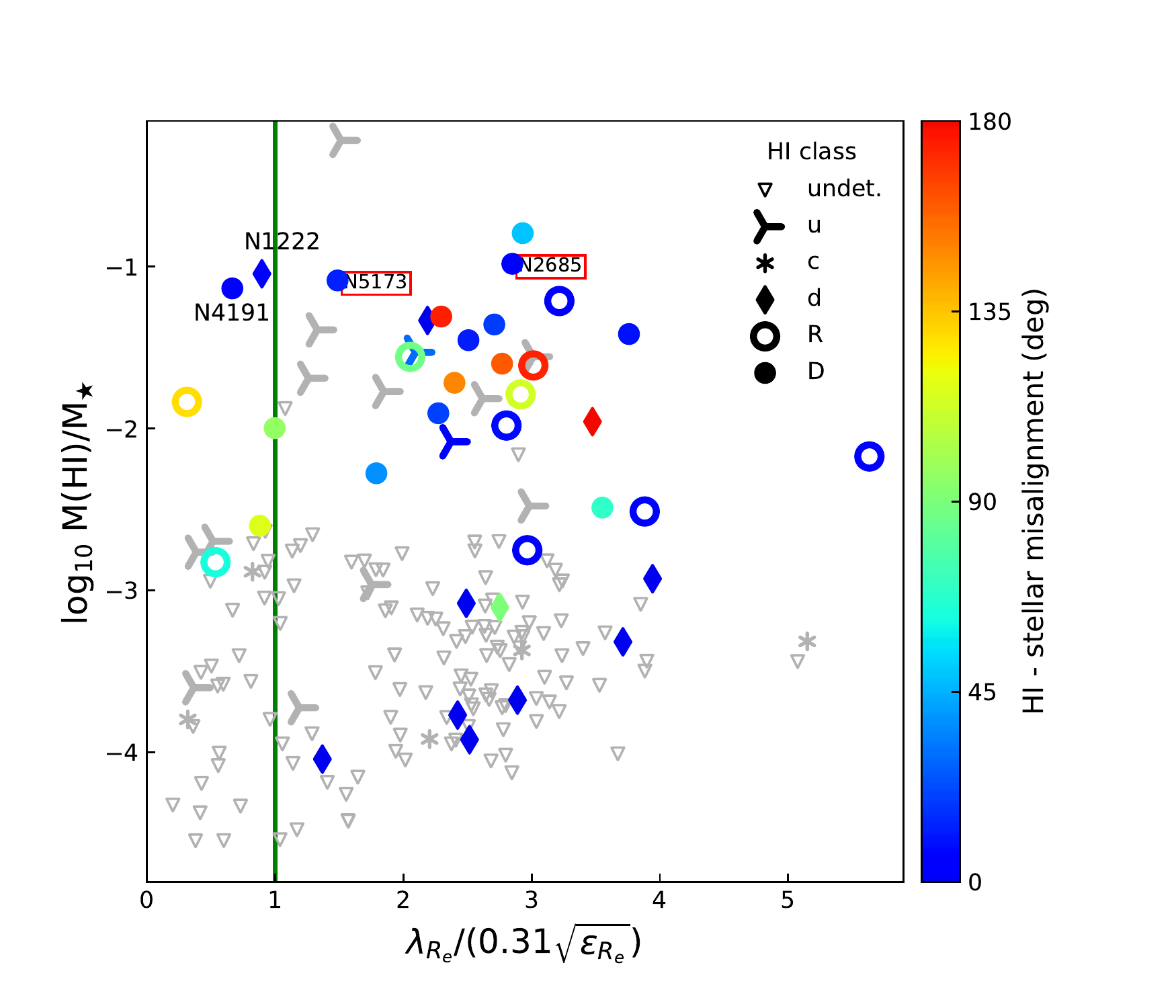}
\caption{An updated version of Figure 5 from \citetalias{serra14}.  \hi\ masses in the \atlas\ sample are plotted
against a scaled version of the stellar specific angular momentum parameter $\lambda_{\mathrm
R_e}$.  Slow rotators are located to the left of the vertical line and fast rotators are to
the right.  The symbol shapes indicate the \hi\ morphology: $u$ are unsettled distributions like tails
or streams, $c$ are small clouds, $d$ are small discs, $D$ are large discs, and $R$ are rings.
The symbol colours indicate the kinematic
misalignments between the \hi\ and the stars, as in \citetalias{serra14}.  Gray symbols mean that
\hi\ is not detected or that it is not settled enough for the misalignment angle to be meaningful.
NGC~5173 and NGC~2685 are also marked because
their dramatic kinematic twists are not adequately captured by the median misalignment value.}
\label{fig:s14fig5update}
\end{figure}

There are several other ways in which the new data emphasize the similarities in the \hi\ properties of 
fast and slow rotators.  
For example, \citetalias{serra14} found that there were no small \hi\ discs (\hi\ class {\it d}, for
discs of similar size
to the stellar body) among slow rotators.  NGC~1222 now fills this niche (Section \ref{sec:1222}). 
The incidence of relaxed \hi\ discs in slow rotators outside of the Virgo Cluster is now 6/26 or 0.23
$\pm$ 0.09, in good agreement with the analogous rate 0.25 $\pm$ 0.05 for fast rotators.
The data in \citetalias{serra14} also suggested a dearth of slow rotators at the highest \hi\ masses; 
specifically, Figures 2 and 5 in \citetalias{serra14} show several fast rotators with HI/stellar mass ratios 
log M(\hi)/M$_\star \approx -1$, but
no such gas-rich slow rotators.
NGC~1222 and NGC~4191 have stellar masses log M$_\star$ = 10.50 and 10.70, respectively
\citep{cap:a3dJAM}, and 
their \hi\ masses give them log M(\hi)/M$_\star = -1.04$ and $-1.13.$
These values fill in the apparent gap in those figures so that slow rotators reach the same high
levels of \hi\ richness as fast rotators (Figure~\ref{fig:s14fig5update}).
The new \hi\ detections are also found in low density environments, as well \citep[several local density
estimates are given in][]{cap-density}, consistent with the trend noted in \citetalias{serra12}.

There is one other facet of the new data that provides a perspective beyond the results of
\citetalias{serra14},
and that is the \hi\ kinematics and its alignment with respect to the stars.
\citetalias{serra14} focused on those alignments as indicators of the evolutionary connections (or the lack thereof)
between gas discs and stellar discs.  Very simply, if a stellar disc forms out of a gas disc, the systems
should share kinematics.  But unlike spirals, early-type galaxies frequently show strong misalignments
between their gas and stellar systems.  \citet{davis_pv} discuss misalignments between stars and 
molecular/ionized gas, and \citet{lagos15} provide some possible interpretations of those misalignments.
Molecular gas is very rare in slow rotators, though \citep{a3dco}, so the \hi\ misalignments are more
interesting for slow rotators, at least in the cases where the \hi\ is relaxed into a regular disc.
\citetalias{serra14} found only two cases of slow rotators with both well-defined \hi\ and stellar kinematics, and in both
of these cases the \hi\ is roughly polar (90\arcdeg\ misalignment).\footnote{These are NGC~3414 and
NGC~5631.}  There are two other cases with well-defined \hi\ kinematics but poorly constrained stellar
kinematics, and those are also polar if one assumes the objects are oblate so that the photometric axis
is a proxy for the kinematic axis.\footnote{These are NGC~3522 and UGC~03960.}
Thus, the evidence from \citetalias{serra14} suggested that \hi\ in slow rotators is most commonly in a polar
configuration, when it is sufficiently relaxed to be in a long-lived configuration.

In contrast, in NGC~1222 and NGC~4191 the \hi\ and stars rotate about nearly identical
axes.
They still have interesting structural features, though; NGC~1222 (Section \ref{sec:1222}) exhibits
prolate-like rotation in its stellar body, so while its gas kinematics are closely aligned with its
stellar kinematics, it may have an uncommon shape.
NGC~4191 (Section \ref{sec:4191}) takes the more common oblate shape, but it is 
also one of the unusual ``2$\sigma$'' galaxies consisting of two coplanar but
counterrotating stellar discs \citep{davor}.  Therefore, its \hi\ corotates with one of the stellar
discs
but counterrotates relative to the other. 
In short, the new data present examples of \hi-rich slow rotators with aligned stellar and \hi\ kinematics.
Figure~\ref{fig:s14fig5update} shows how these two compare to the other slow and fast rotators in
terms of their \hi\ masses, kinematics and morphologies; it reinforces the similarities between the
\hi\ properties of slow rotators and fast rotators.

The simulations of \citetalias{serra14} and \citet{lagos15} have convincingly shown that gas kinematics in
early-type galaxies are sensitive indicators of the galaxies' evolutionary histories, so improved
simulations should attempt to reproduce the full spectrum of shapes, structures, and misalignments
exhibited by early-type galaxies.

\begin{table*}
\caption{Updated data on \hi\ in \atlas\ slow rotators.}
\label{tab:SRstats}
\begin{tabular}{lclcccccclc}
\hline
Name & Slow/Fast & Kin.~Type & class & core? & log M(\hi) & log M(\hi)/M$_\star$ &
\hi\ morph & Notes & source & Virgo \\
 & $R_e$; $R_e/2$ &          &        &       & (\solmass) &      &   & &  & Cluster \\
(1) & (2) & (3) & (4) & (5) & (6) & (7) & (8) & (9) & (10) & (11) \\
\hline
NGC0661    &  S S  & NRR/CRC    & c &   \nodata   &  $<$ 7.37 &  $< - 3.56$ &  \nodata  &  \nodata  &       S12  &   \\
NGC1222    &  S S  & NRR/NF     & b &   \nodata   &  =   9.46 &  $= - 1.04$ &         d &  \nodata  &       *    &   \\
NGC1289    &  S S  & NRR/CRC    & c &   \nodata   &  $<$ 7.67 &  $< - 3.05$ &  \nodata  &  \nodata  &       *    &   \\
NGC3414    &  S S  & NRR/CRC    & c & $\setminus$ &  =   8.28 &  $= - 2.83$ &         D &     P,R,L &       S12  &   \\
NGC3522    &  S S  & NRR/KDC    & c &   \nodata   &  =   8.47 &  $= - 1.83$ &         D &       P,R &       S12  &   \\
NGC3608    &  S S  & NRR/CRC    & c &      $\cap$ &  =   7.16 &  $= - 3.80$ &         c &  \nodata  &       S12  &   \\
NGC3796    &  S S  & NRR/2s     & d & $\setminus$ &  $<$ 7.10 &  $< - 2.89$ &  \nodata  &  \nodata  &       S12  &   \\
NGC4168    &  S S  & NRR/KDC    & c &      $\cap$ &  $<$ 7.46 &  $< - 3.84$ &  \nodata  &  \nodata  &       S12  &   \\
NGC4191    &  S S  & NRR/2s     & d &   \nodata   &  =   9.57 &  $= - 1.13$ &         D &       W,L &       *    &   \\
NGC4261    &  S S  & NRR/NF     & b &      $\cap$ &  \nodata   &  \nodata   &  \nodata  &  \nodata  &  \nodata  &   \\
NGC4365    &  S S  & NRR/KDC    & c &      $\cap$ &  $<$ 7.52 &  $< - 4.01$ &  \nodata  &  \nodata  &  $\alpha$ &   \\
NGC4374    &  S S  & NRR/LV     & a &      $\cap$ &  $<$ 7.26 &  $< - 4.33$ &  \nodata  &  \nodata  &       S12  &  V  \\
NGC4406    &  S S  & NRR/KDC    & c &      $\cap$ &  =   8.00 &  $= - 3.60$ &         u &  \nodata  &       S12  &  V  \\
NGC4458    &  S S  & NRR/KDC    & c & $   \wedge$ &  $<$ 6.91 &  $< - 3.12$ &  \nodata  &  \nodata  &       S12  &  V  \\
NGC4472    &  S S  & NRR/CRC    & c &      $\cap$ &  $<$ 7.23 &  $< - 4.54$ &  \nodata  &  \nodata  &  $\alpha$ &  V  \\
NGC4476    &  S F  & RR/NF      & e & $   \wedge$ &  $<$ 7.29 &  $< - 2.71$ &  \nodata  &  \nodata  &       L05  &  V  \\
NGC4486    &  S S  & NRR/LV     & a &      $\cap$ &  \nodata   &  \nodata   &  \nodata  &  \nodata  &  \nodata  &  V  \\
NGC4528    &  S F  & NRR/2s     & d & $\setminus$ &  $<$ 7.18 &  $< - 2.94$ &  \nodata  &  \nodata  &  S12/$\alpha$  &  V  \\
NGC4550    &  S S  & NRR/2s     & d & $   \wedge$ &  $<$ 6.89 &  $< - 3.51$ &  \nodata  &  \nodata  &       S12  &  V  \\
NGC4552    &  S S  & NRR/NF     & b &      $\cap$ &  $<$ 6.87 &  $< - 4.33$ &  \nodata  &  \nodata  &       S12  &  V  \\
NGC4636    &  S S  & NRR/LV     & a &      $\cap$ &  $<$ 6.85 &  $< - 4.55$ &  \nodata  &  \nodata  &       *    &   \\
NGC4690    &  S S  & NRR/NF     & b &   \nodata   &  $<$ 7.80 &  $< - 2.82$ &  \nodata  &  \nodata  &       *    &   \\
NGC5198    &  S S  & NRR/NF     & b &      $\cap$ &  =   8.49 &  $= - 2.70$ &         u &  \nodata  &       S12  &   \\
NGC5322    &  S S  & NRR/CRC    & c &      $\cap$ &  $<$ 7.34 &  $< - 4.19$ &  \nodata  &         A &       S12  &   \\
NGC5481    &  S S  & NRR/KDC    & c &   \nodata   &  $<$ 7.21 &  $< - 3.40$ &  \nodata  &  \nodata  &       S12  &   \\
NGC5557    &  S S  & NRR/NF     & b &      $\cap$ &  =   8.57 &  $= - 2.76$ &         u &  \nodata  &       S12  &   \\
NGC5576    &  S S  & NRR/NF     & b & $   \wedge$ &  $<$ 7.30 &  $< - 3.58$ &  \nodata  &  \nodata  &       *    &   \\
NGC5631    &  S F  & NRR/KDC    & c &   \nodata   &  =   8.89 &  $= - 2.00$ &         D &     M,W,L &       S12  &   \\
NGC5813    &  S S  & NRR/KDC    & c &      $\cap$ &  $<$ 7.51 &  $< - 4.08$ &  \nodata  &  \nodata  &       *    &   \\
NGC5831    &  S S  & NRR/KDC    & c & $\setminus$ &  $<$ 7.28 &  $< - 3.59$ &  \nodata  &  \nodata  &       *    &   \\
NGC5846    &  S S  & NRR/LV     & a &      $\cap$ &  $<$ 7.20 &  $< - 4.37$ &  \nodata  &  \nodata  &       *    &   \\
NGC6703    &  S F  & NRR/LV     & a & $\setminus$ &  $<$ 7.18 &  $< - 3.80$ &  \nodata  &  \nodata  &       S12  &   \\
NGC7454    &  S S  & NRR/NF     & b &   \nodata   &  $<$ 7.16 &  $< - 3.47$ &  \nodata  &  \nodata  &       S12  &   \\
PGC028887  &  S F  & NRR/KDC    & c &   \nodata   &  =   7.65 &  $= - 2.88$ &         c &  \nodata  &       S12  &   \\
PGC050395  &  S S  & NRR/CRC    & c &   \nodata   &  $<$ 7.51 &  $< - 2.63$ &  \nodata  &  \nodata  &       S12  &   \\
UGC03960   &  S F  & NRR/NF     & b &   \nodata   &  =   7.79 &  $= - 2.60$ &         D &       W,L &       S12  &   \\

\hline
\end{tabular}

These are the \atlas\ galaxies classified as slow rotators based on stellar kinematics within one effective
radius (R$_e$).  Column 2 gives the fast/slow classifications within R$_e$ and R$_e$/2 from
\citet{emsellem_a3d}.  Columns 3 and 4 give the stellar kinematic types and classes from
\citet{davor}.  
Kinematic type abbreviations are as follows: RR = regular rotator; NRR = non-regular rotator; CRC
= counterrotating core; KDC = kinematically decoupled core; LV = low velocity core; 2s = two velocity
dispersion peaks, symptomatic of counterrotating discs; NF = no other distinguishing features.
The kinematic classes in column 4 are more general.  Class a includes those galaxies
without detected rotation in the field; class b galaxies have complex velocity fields not otherwise
classified. Class c galaxies have kinematically distinct cores, d galaxies have two counterrotating
discs, and e galaxies have simple aligned rotation.
Column 5 gives the nuclear structure classification from \citet{davor_cusp}, as $\cap$ = core, $\wedge$ = intermediate,
and $\setminus$ = power law.
Columns 6 to 9 give the \hi\ mass, \hi\ to stellar mass ratio, morphology and kinematic notes.
Stellar masses are taken from \citet{cap:a3dJAM}.  
From \citet{serra12}, the \hi\ morphological classes in column 8 are as follows: u = unsettled, such as tails
or streams; c = clouds, small and irregular; d = small disc; D = large disc.
Further \hi\ kinematic notes in column 9 include these:
A = \hi\ detected (also) in absorption; 
L = lopsided; 
M = \hi\ misaligned relative to the stellar kinematics; P = polar orientation ($\approx$ 90\arcdeg\
misalignment); R = ring; W = warp.
The \hi\ reference is in column 10, with a key as follows.
S12 = \citet{serra12}; * = this paper; L05 = \citet{lucero4476}; $\alpha$ = the ALFALFA survey
\citep{giovanelli05}, and R.\ Giovanelli, private communication.
Column 11 indicates Virgo Cluster membership according to \citet{cappellari_a3d1}.
\end{table*}

\subsection{Discussion: Implications for the formation of slow rotators}

\citet{cappellari_araa} has summarized the formation processes of slow rotators as falling in 
two broad categories, one of which takes place
in clusters and groups (dense environments) and the other of which takes place in the field. 
The first category includes scenarios such as a brightest cluster galaxy disrupting and accreting other
cluster members, and after many such mergers, a relaxed cluster can contain one enormous slow rotator 
that is well-separated from other cluster members in the mass-size plane.
In less relaxed clusters, where substructure is present, there may be multiple slow rotators associated
with the sub-groups; however, these should eventually merge as well.
In the field, however, the formation of slow rotators is expected to be more of a stochastic process which happens
when the orbital configuration of a major merger is ``just right" to produce a remnant with very low
specific angular momentum.  This latter process can also produce relatively small, low-mass slow
rotators, and it can produce dispersion-dominated galaxies as well as rotation-dominated galaxies
whose {\it net} angular momentum is low because they contain two counter-rotating structures.

In this context, the fact that the \hi\ content of slow rotators matches that of fast rotators is
probably related to the overall \hi\ content of the universe at low redshift.  Both fast and slow
rotators can form in mergers, some of which may be gas-rich, and both can accrete cold gas from their
surroundings, if there is any cold gas in the vicinity.
Of course there is still a very strong dependence of \hi\ content on the local galaxy density, as noted
by \citetalias{serra12} and confirmed here.  Thus we might not expect the most massive slow rotators in galaxy clusters
to have HI.

Additional evidence for a variety of formation paths for early-type galaxies comes from their nuclear
structures:  specifically, the presence or absence of inner cores in the surface brightness profiles.
Scouring by an inspiraling supermassive black hole binary is a popular model for the formation of
cores, though not the only model \citep{davor_cusp}.
It is also generally expected that it should be much more difficult to form or maintain a core when
cold gas is present.  The dissipational nature of the cold gas should make it sink towards the
nucleus and form stars, refilling the scoured core.
And as discussed by \citet{davor_cusp}, cored galaxies are commonly slow rotators and core-less
galaxies are commonly fast rotators, though there are exceptions to those rules.

In this context, with our updated statistics on the \hi\ content of slow rotators, we
reassess possible links between their cold gas content and their nuclear structure.
Table~\ref{coretable} presents data on \hi\ and CO detections in the \atlas\ galaxies with nuclear
surface brightness classifications.
We first note that detections of \hi\ emission are slightly more common in cored than in 
core-less slow rotators; \hi\ is detected in 4 of 13 cored slow rotators and one of nine core-less slow
rotators.  However, it is useful to remember that much of that \hi\ emission is in the outskirts of the
galaxies.  Gas in the ``c" or ``u" morphologies, i.e. irregular clouds or unsettled distributions, may
never have been close enough to the nucleus to refill a scoured core.  
Thus, we also tabulate detections of {\it central} \hi\ or CO emission on kpc scales, where we find cold
gas present in 3 of 9 core-less slow rotators but in only one of the 13 cored slow rotators.

This relatively simple association suffers, of course, from small number statistics, but the picture is
also complicated by the fact that the data are inhomogenous in quality
and several of the cored slow rotators are also detected in other species.
The CO detection in NGC~4374 corresponds to (6$\pm$1)\e{6} \solmass\ of cold molecular gas
\citep{boizelle}, and we
should note that it comes from ALMA data of much higher sensitivity and resolution than that 
available for any other galaxy in the table.
Furthermore, NGC~4486, NGC~4636, and NGC~5813 are detected in the \oi\ 63 $\micron$ line, in their central
kpc or so, and this line is usually associated with the outer envelopes of molecular clouds (Section
\ref{sec:nondets}).
NGC~4261 and NGC~4374 are also detected in the mid-IR lines of excited H$_2$ \citep{ogle2010}, giving masses
in the range of $10^5$ to $10^7$ \solmass\ of molecular gas at temperatures of a few hundred K.
This CO, H$_2$ and neutral O could represent gas that has cycled through an AGN feedback episode of
the type described by \citet{russell2017}.
The FIR and mid-IR spectroscopy data are only available for a few early-type galaxies, so it is
difficult to make broader conclusions about the incidence of these species, and it is also 
not clear whether these detections represent gas that is capable of star formation (filling in the
cores).

In short, cored galaxies are detected in \hi\ emission on large scales at a similar or higher rate as
core-less galaxies; they are not commonly detected in \hi\ emission in their nuclei, but they are
occasionally detected in other species associated with cold atomic and/or molecular gas.
In any case it is clear that the issue of star-forming gas and stellar cores requires more work.

\begin{table}
\centering
\caption{Cold gas and cores in slow rotators.}
\label{coretable}
\begin{tabular}{l|c|c|c|c}
 & \hi & no \hi & Total & central \hi \\
 &     &        &       & or CO  \\
\hline
Core-     & 1 & 8 & 9 & 3 \\
less      & 3414 (D) & 3796, 4458, 4476, & & 3414, \\
          &      & 4528, 4550, 5576, & & 4476,\\
          &      & 5831, 6703 & & 4550\\
\hline
Core      & 4 & 9 & 13 & 1 \\
          & 3608 (c), 4406 (u), & 4168, 4365, 4374, & & 4374\\
          & 5198 (u), 5557 (u)  & 4472, 4552, 4636, & & \\
          &                     & 5322, 5813, 5846 & & \\
\hline
\end{tabular}
The table lists the number of galaxies in each category and their NGC identifiers.
For \hi-detected galaxies, the \hi\ morphological classification from \citetalias{serra12} is also given.
This table includes only the ones with {\it both} published 
core/core-less status \citep{davor_cusp} {\it and} good \hi\ and CO data.  Thus NGC~4486
and NGC~4261, both cored galaxies, are not listed because they lack good \hi\ data.  
Central \hi\ data are taken from \citet{a3d_cmd} and CO data
from \citet{a3dco} and \citet{crocker4550}, with a recent CO detection of NGC~4374 from \citet{boizelle}.
\end{table}

\section{Notes on the \hi\ non-detections}\label{sec:nondets}

Broadly speaking, the \hi\ detection rate in early-type galaxies is only 32\% \citepalias{serra12}, so the many
non-detections in the current set is not surprising.  But in some cases we have additional information
about other phases of the interstellar medium and the environment of our targets, and this information
can help place the non-detections into a broader context.

For comparison to the \hi\ upper limits, we note that FIR photometry from the Herschel Reference Survey
gives estimated dust masses for two of these galaxies.  \citet{mwlsmith2012} find log M$_d$ (\solmass) = 5.06 $\pm$ 0.19 for NGC~4636 and $< 5.51$ for
NGC~5576.  Adopting a M(\hi)/M$_d$ ratio of 100, based on empirical correlations from the other members
of the Herschel Reference Survey \citep{cortese2016}, the dust limit for NGC~5576 is consistent with
our \hi\ nondetection.  The nominal M(\hi)/M$_d$ ratio suggests that we should have detected $10^7$
\solmass\ of \hi\ in NGC~4636, but that is not a compelling inconsistency given the large range in
observed M(\hi)/M$_d$ ratios. 
The other HRS galaxies exhibit values from
10 to 1000, and two other \atlas\ slow rotators (NGC~3414 and NGC~4406) have M(\hi)/M$_d$ ratios of 100
and 23, respectively.  Thus it would not be unusual to have M(\hi)/M$_d < 100$ in NGC~4636 and the HI
nondetection could still be consistent with the dust detection.

Besides dust masses, we note that NGC~4486, NGC~4472, NGC~4636, NGC~5813, and NGC~5846 are 
detected in \cii\ 158$\mu$m emission \citep{brauher,werner}.
In nearby late-type galaxies, \cii\ emission is usually associated with cold molecular gas as it
should form a photodissociated envelope on the UV-exposed skin of a molecular cloud.  But of course it
can also arise in ionized gas, and in fact \citet{wilson4125} have suggested that the \cii\ emission in
the early-type galaxy NGC~4125 should be entirely attributed to ionized gas.  In this context it is not
obvious whether the \cii\ detections mentioned above should lead us to expect \hi\ or CO detections.
But there are also \oi\ 63$\mu$m detections in NGC~4486, NGC~4636, and NGC~5813, and these are
unambiguously associated with neutral atomic or molecular gas.  Further quantitative analysis should be
carried out to check whether the simultaneous detections of \oi\ and the non-detections of \hi\ in
NGC~4636 and NGC~5813 are significant.  The \oi\ detections do serve as a reminder that significant
quantities of cold gas can be hiding beneath the present sensitivity limits.

\begin{figure*}
\includegraphics[scale=1.1,trim=2.5cm 9cm 1cm 3cm,clip]{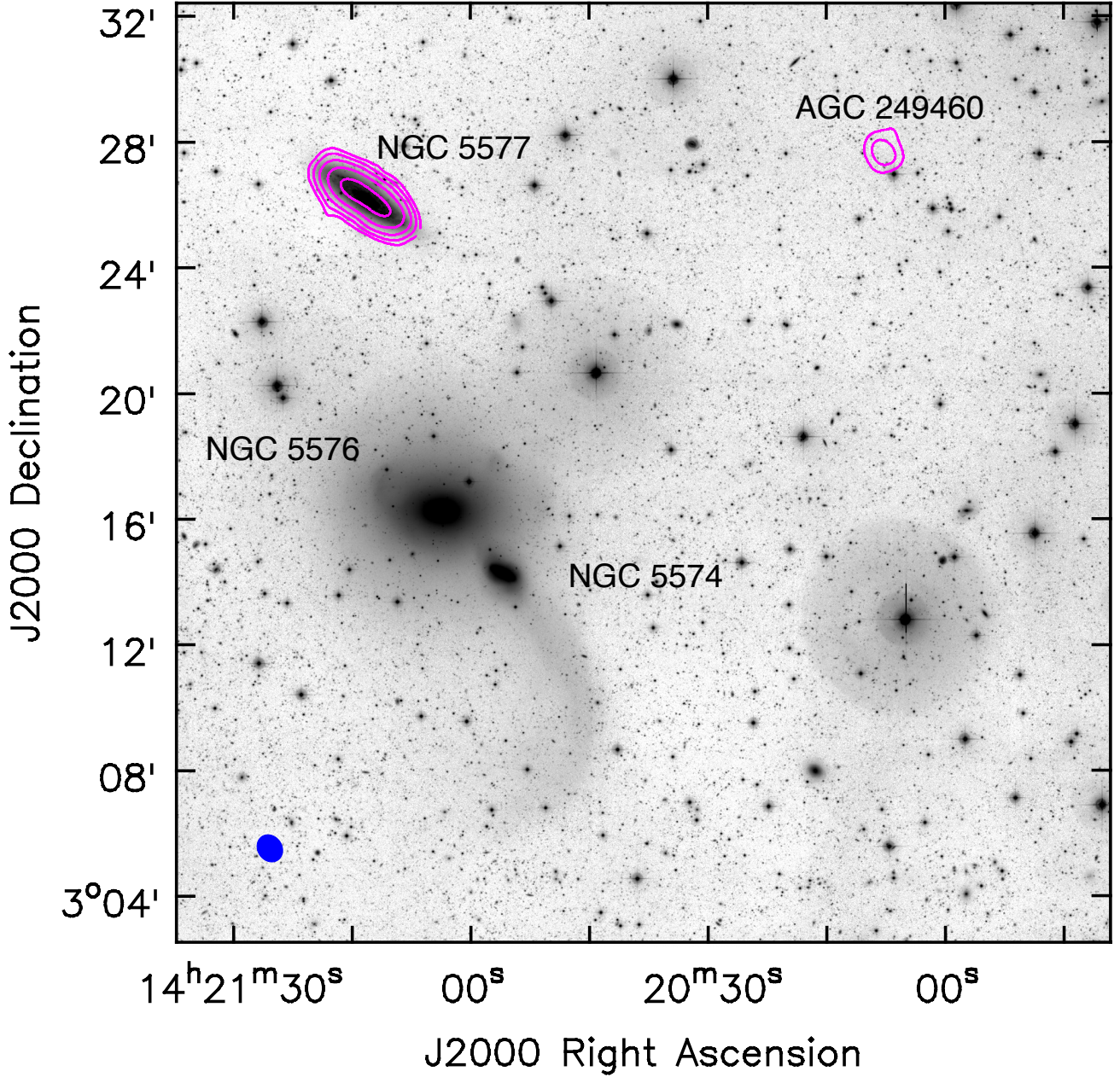}
\caption{The NGC~5577, NGC~5576, and NGC~5574 group. 
The optical image is the MATLAS $g$ data; the \hi\ column density image is overlaid, and contours are $(3, 6, 12, 24, 48) \times 2.3\times 10^{19}$
\persqcm.  
A primary beam correction has been applied to the \hi\ column densities.  The blue ellipse in the bottom
left corner indicates the \hi\ resolution.  AGC~249460 is discussed in the Appendix.}
\label{fig:5577m0}
\end{figure*}

Finally, it is notable that NGC~5576 and NGC~5574 are undergoing a strong interaction, and NGC~5574
has developed impressive tidal tails (Figure~\ref{fig:5577m0}).
Evidently NGC~5574 has been relatively poor in cold gas for some time, as there is no stripped \hi\ in the
tails.
Strong \hi\ emission from the spiral NGC~5577 could cause confusion in single dish HI
observations of the region, though.
The system is discussed in more detail in the Appendix. 

\section{Cold gas in the merger remnant NGC~1222}\label{sec:1222}

\begin{figure*}
\hbox{
\includegraphics[scale=0.7,trim=2cm 4cm 5.7cm 5cm,clip]{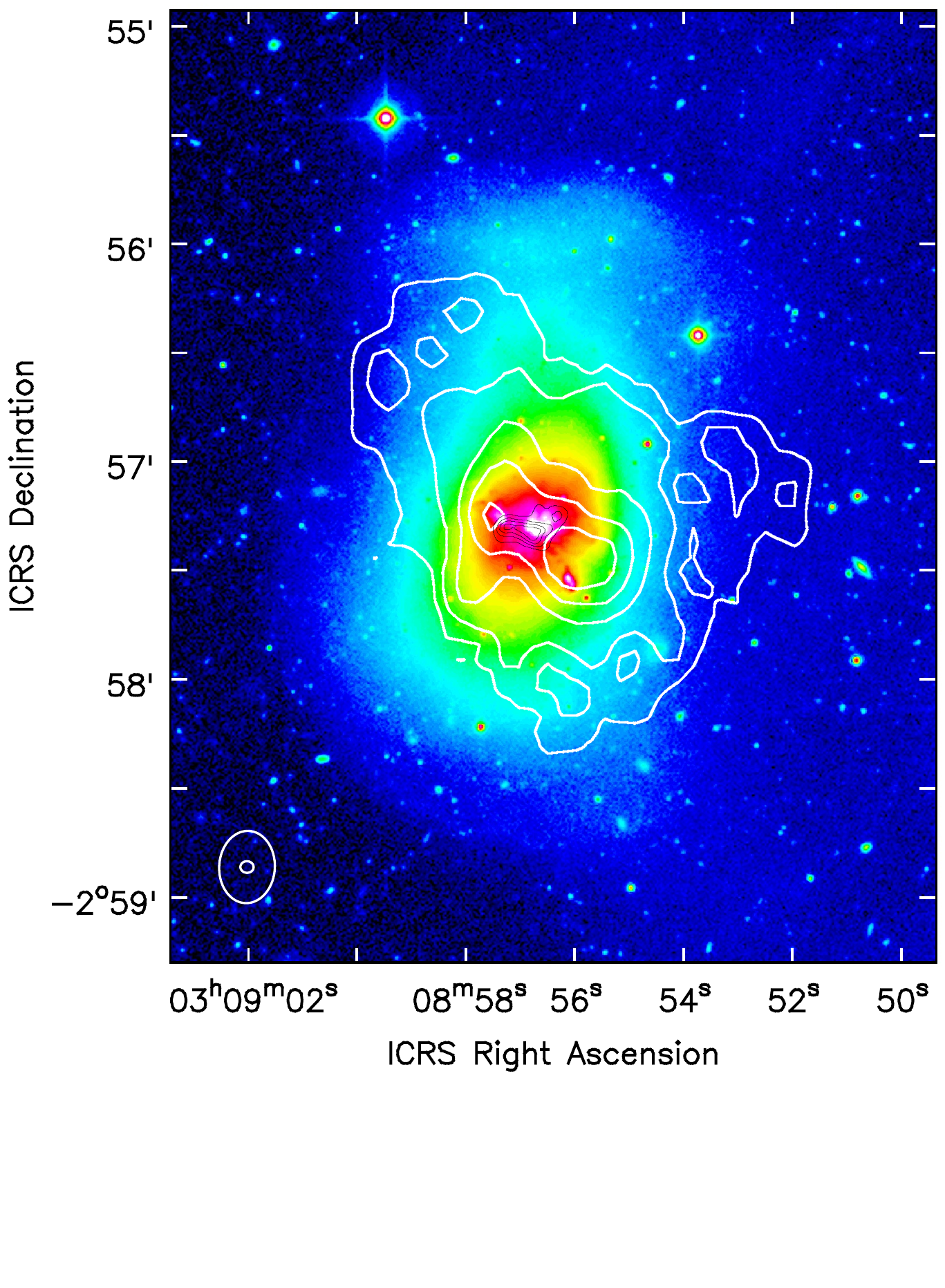}
\vbox{
\hbox{
\includegraphics[scale=0.25,trim=1cm 8cm 3cm 0cm, clip]{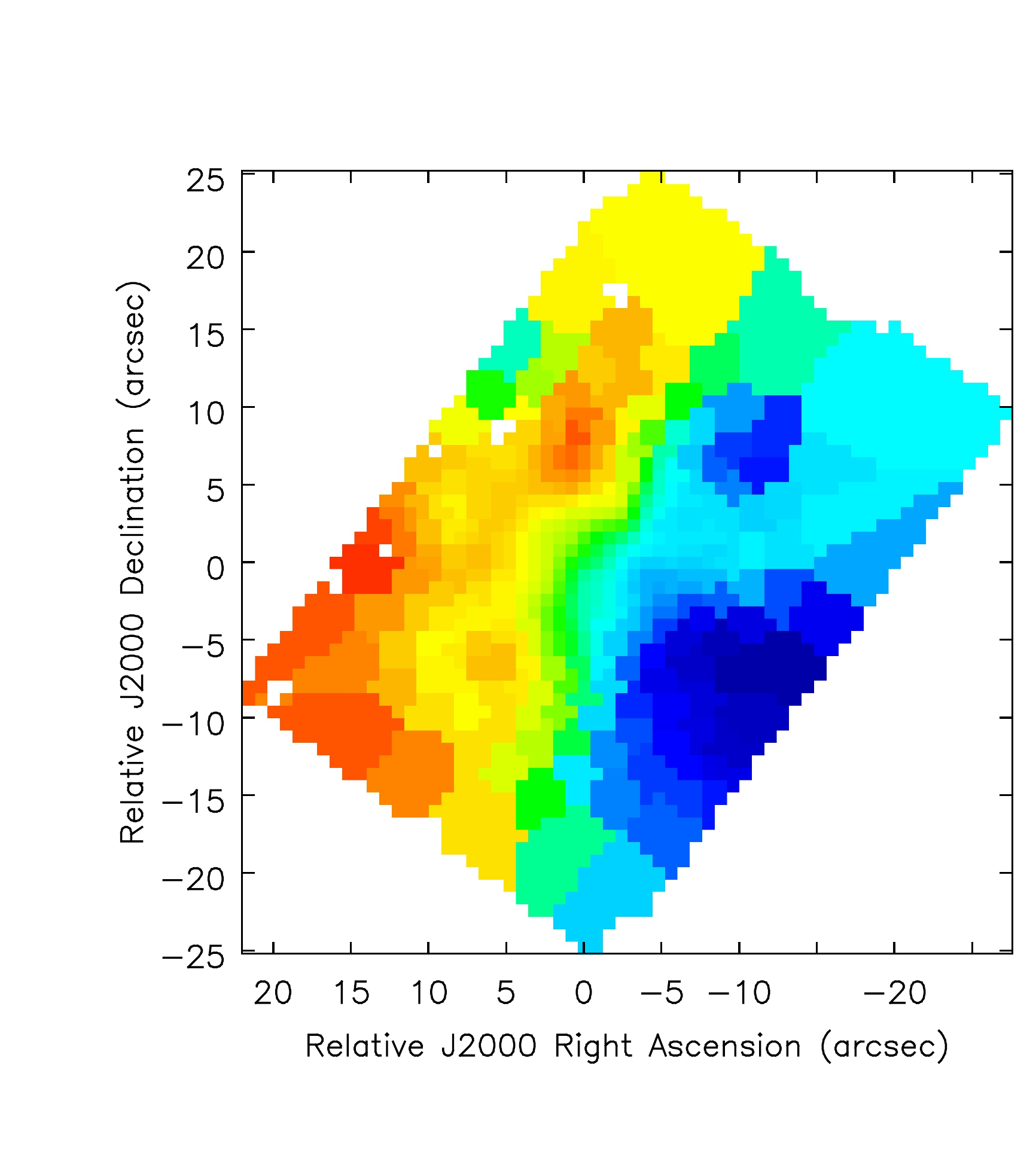}
\includegraphics[scale=0.25,trim=3.6cm 8cm 2cm 0cm, clip]{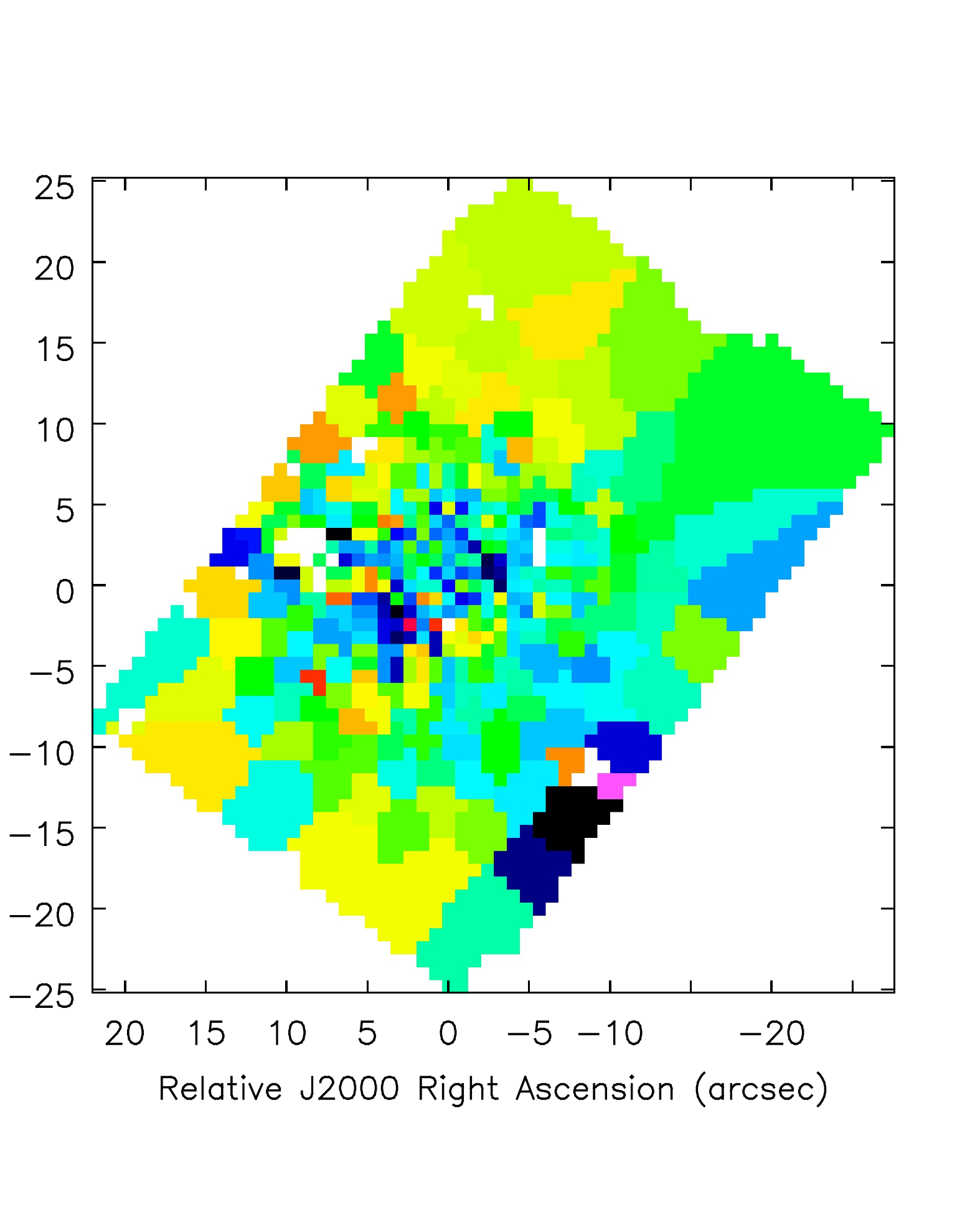}
}
\includegraphics[scale=0.45, trim=22.5cm 6.5cm 3cm 2cm]{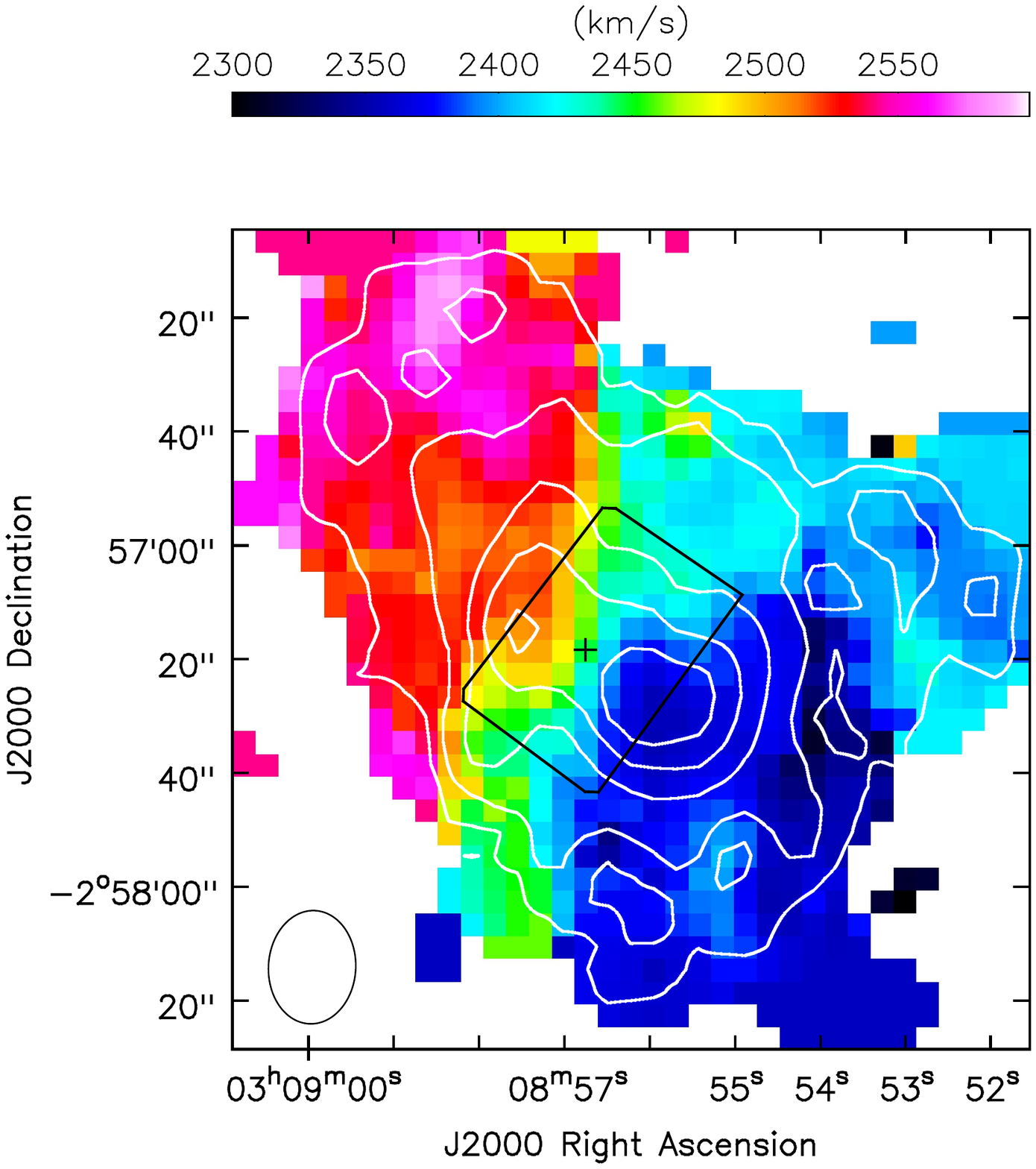}
}
}
\caption{NGC~1222.  Left: Optical is the g-band MATLAS image from \citet{duc15}.  White
contours show the \hi\ distribution; contours are (0.1, 0.2, 0.4, 0.6, 0.8) times the peak, which is 1.14
\jybkms\ or 4.1\e{21}\persqcm.  Black contours show the CO distribution from \citet{a3dcocarma}; they are
(0.2, 0.4, 0.6, 0.8) times the peak value of 17.5 \jybkms\ or 4.2\e{22} molecules\persqcm\ at a
conversion factor of 3\e{20} molecules\persqcm\ per (K~\kms).
Right: Velocity fields in NGC~1222.  The lower image shows \hi\ velocities; the black cross marks the
optical nucleus and the black polygon shows the field of view for the optical velocity data.  The upper
pair are the \oiii\ velocities (left) and the stellar velocities (right).  All the velocity fields use
the same colour scale.}
\label{fig:1222m0}
\end{figure*}

NGC~1222 is a well-known recent merger remnant, currently undergoing starburst activity.  New, deep optical observations from \citet{duc15} clearly show 
several shells and tails (Figure~\ref{fig:1222m0}).
It is evidently in a merger-driven star formation episode, and given its star formation
rate \citep{davis14} its gas consumption time is 1 to 2 Gyr.  On that kind of a timescale we would thus
expect the optical features to fade and the cold gas content to decline as it approaches the structure
and appearance
of the other slow rotators.  At present, though, it is unique among the slow rotators in being the only
one detected in both \hi\ and CO emission; its HI/H$_2$ mass ratio of 2.4 $\pm$ 0.4 is similar to typical
values found in fast rotators, whereas slow rotators tend to be H$_2$-poor \citep{a3d_cmd}.
The \hi\ and CO emission are coincident with a dust lane, 
with the CO emission nestled into a central depression in the \hi\ column density,
and the gas and dust are all oriented roughly perpendicular to the stellar photometric major axis.
The previous literature on the galaxy has not discussed the \hi\ kinematics and its relation to the merger
features, so we present that information here.

\subsection{Prolate-like rotation in gas and stars}

In more detail, the structure of the galaxy can be quantified by applying the kinemetry
routines of \citet{kinemetry} to the deep $g$ image and velocity fields.
Regions interior to a radius of ~30\arcsec\ in the optical image are strongly
disturbed by dust and areas of recent star formation activity.  Exterior to that, the photometric
position angle shows a gentle twist from $-$60\arcdeg\ at 30\arcsec\ to
$-$10\arcdeg\ at radii of 200\arcsec.  The outer isophotes are again strongly distorted by the shells
and tails.  For comparison, within the SAURON field of view the stellar kinematic PA is found to be
43.0\arcdeg\ $\pm$ 9.2\arcdeg\ \citep{davor}, so that the stellar rotation is nearly about the
photometric major axis.  This long-axis or prolate-like rotation suggests that the stellar body of the 
galaxy may have a prolate or triaxial shape.

The gas kinematics are also roughly aligned with the stellar kinematics
(Figure~\ref{fig:1222m0}).
We estimate the kinematic PA of the ionized gas to be 63\arcdeg\ $\pm$ 11\arcdeg; 
that of the \hi\ to be 48.5\arcdeg $\pm$ 1.3\arcdeg;
and that of the CO to be 40\arcdeg\ $\pm$ 2\arcdeg.  
The systematic trend in position angle, from ionized gas through \hi\ to CO, is suggestive that the
gas (having been disturbed in the merger) is settling towards a stable plane as it cools.
On the other hand it is notable that both the \hi\ and
CO are lumpy, so the true uncertainties on the PAs may be larger than the values
formally suggested by the statistical analysis.
In short, in NGC~1222 the stellar rotation is prolate-like, and all the gas kinematic PAs are 
a close match to the
stellar kinematic PA.  If the galaxy is indeed prolate then the gas is close to its equatorial
plane.\footnote{In this case, despite the superficial similarities, the structure of NGC~1222 
is different from that of NGC~5128 (Cen~A).  \citet{sparke_cena} shows that NGC~5128 appears to be an
oblate galaxy with a polar disc.}
After it has aged and reddened, NGC~1222 may look like NGC~5485, another \atlas\ galaxy with
prolate-like rotation and a dust disc along its photometric minor axis (kinematic major axis).
Of course, a recent merger like NGC~1222 may not yet have approached its final stable configuration, so it
is difficult to infer its true shape.  Stellar kinematics at larger radii would also be helpful for
determining whether the prolate-like rotation persists farther out.

Kinemetric analysis of the \hi\ velocity field also indicates that at moderate radii (10\arcsec\ to
40\arcsec), where the velocity field is well behaved, the inclination of the \hi\ disc is
38\arcdeg\ $\pm$ 6\arcdeg.
With this inclination, the inferred rotation speed in the \hi\ disc is 162 \kms\ $\pm$ 20 \kms. 
The dynamical mass interior to 40\arcsec\ (6.5 kpc) is then 4.0\e{10} \solmass, which is consistent
with the dynamical mass of 3.2\e{10} \solmass\ inferred by \citet{cap:a3dJAM} from stellar kinematics.  Given the irregular HI
distribution at larger radii, dynamical masses might not be reliable further out; in any case, there
is no strong evidence of dynamical inconsistency.

\subsection{Tails, shells, and the formation of NGC~1222}

Like NGC~1222, many of the \atlas\ galaxies exhibit stellar shells and tails at low surface brightness;
two-thirds of the ones imaged deeply by \citet{duc15} have such features.
From this perspective NGC~1222 may be something like a younger version of NGC~5557,
which was studied in more detail by 
\citet{p9-5557}.  In NGC~5557, diffuse \hi\ emission is clearly associated with a remarkable long stellar
tail, and the galaxy is interpreted as the remnant of a major gas-rich merger.  That merger probably drove a
great deal of cold gas to the center of the galaxy, but the central gas has since been consumed, and
\hi\ only remains out in the stellar tail where the dynamical timescales are long.
In contrast to NGC~5557, we detect no \hi\ emission associated with the stellar
tails and shells in NGC~1222, though it should be noted that the sensitivity of these VLA data 
is not as good as the WSRT data for NGC~5557.  At the sensitivity of the NGC~1222 data, we would have
detected \hi\ like that in the shells of NGC~5128 and similar shell
galaxies \citep{schiminovich_cena, schiminovich_2865, schiminovich2013}, 
but we would not have detected the faint clouds near NGC~5557.
Thus, since the \hi\ and CO in NGC~1222 are not obviously aligned with or associated with the stellar tails and 
shells, it's not clear whether the cold gas took part in the event that formed the 
stellar features.  It could have been accreted afterwards.  Deeper \hi\ observations and more careful
analysis of the stellar populations in the shells could help to clarify the nature of the recent
merger that formed NGC~1222.

In this regard, NGC~1222 is also interesting for its comparison to numerical simulations of galaxy
formation.
\citet{li2017} have studied the merger histories of prolate galaxies in the Illustris
simulation, and they find 
their prolate galaxies were primarily formed in a major dry merger with a radial geometry.
\citet{tsatsi2017} also describe a merger that forms a prolate galaxy, but again it is a dry merger.
If these kinds of collisionless merger scenarios describe the history of NGC~1222, then the cold gas presumably was accreted after the
merger.
Strictly speaking we should note that these analyses focus on galaxies with stellar masses
M$_\star > 10^{11}$ \solmass, but NGC~1222 (M$_\star \sim$ 3.2\e{10} \solmass) is not far below the
range studied.
On the other hand,
\citet{li2017} also find two cases of prolate galaxies formed through mergers where one progenitor 
had 10--15\% of its baryonic mass in the form of gas.  
\citet{ebrova} identify several other cases in the Illustris simulation in which
gas-rich mergers formed massive galaxies with prolate-like rotation.
As NGC~1222
currently has M(HI+H$_2$)/M$_\star \sim 0.13,$ this gas-rich merger scenario might also be roughly appropriate
for it.

\section{An Unusually Large \hi\ disc in NGC~4191}\label{sec:4191}

Amongst the \atlas\ slow rotators in Table~\ref{tab:SRstats}, NGC~4191 is unusual for having the
largest \hi\ mass and disc diameter.  Its \hi\ mass and diameter are, in fact, in
league with the properties of nearby spirals.  It was classified by \citet{davor} as a ``two
$\sigma$-peak'' galaxy, meaning that it consists of two superposed, dynamically cold, counterrotating
stellar discs, and these galaxies are often found to be slow rotators because the two discs'
rotations interfere destructively in the mean velocity field.  Its rotational kinematics were also
studied in more detail by \citet{coccato}.  Despite these interesting kinematics, a deep optical image \citep{duc15} shows little evidence
for extended, low-surface brightness features like tails or shells.
Its unusual \hi\ disc, in combination with its counterrotating stellar discs, make it particularly
interesting for studies of the gas kinematics in comparison to the stellar kinematics.

\subsection{\hi\ distribution and kinematics}

\begin{figure*}
\hbox{
\includegraphics[scale=0.55, trim=2.5cm 4.5cm 4.7cm 2cm,clip]{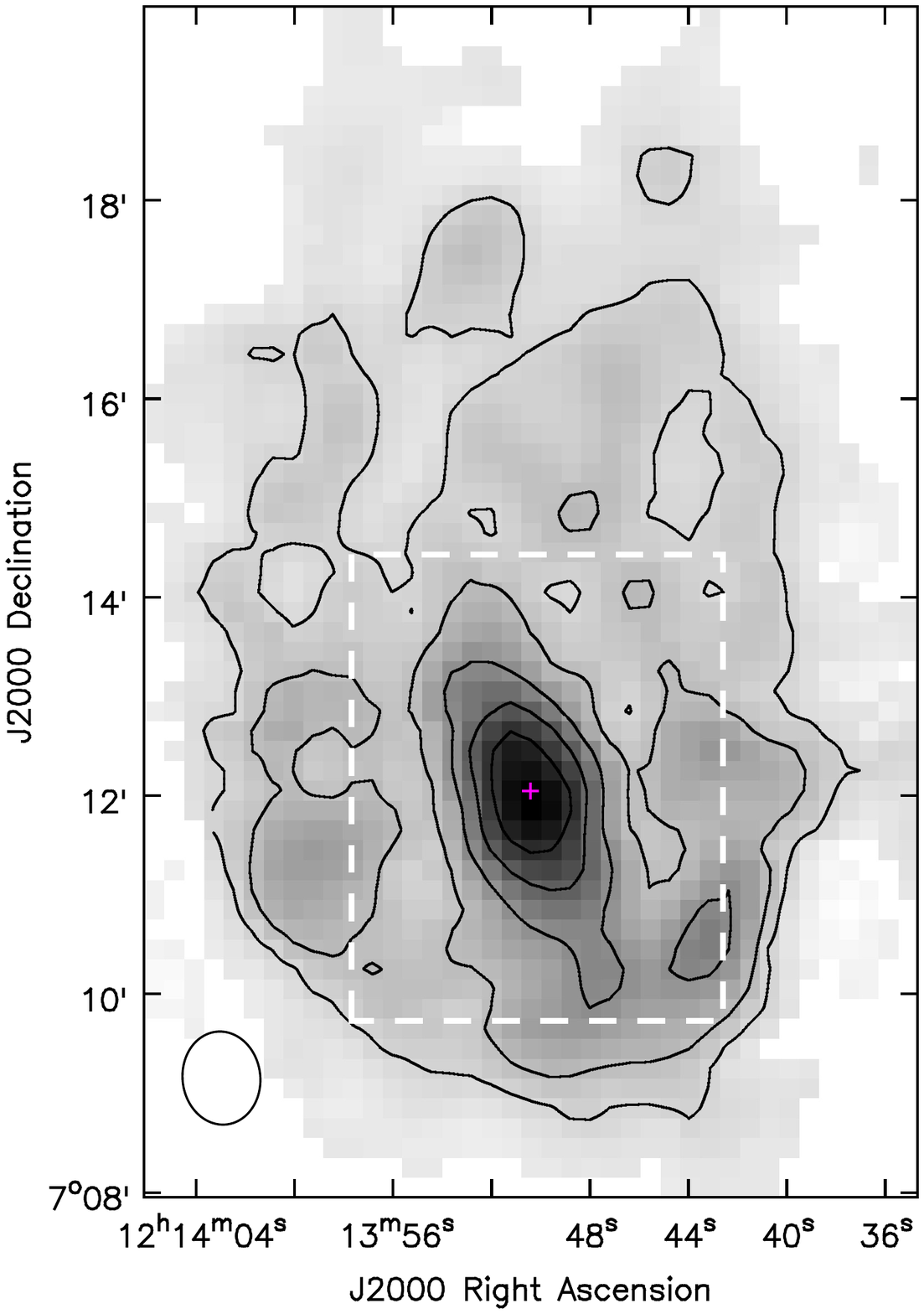}
\includegraphics[scale=0.65, trim=3cm 8.5cm 1.5cm 2cm,clip]{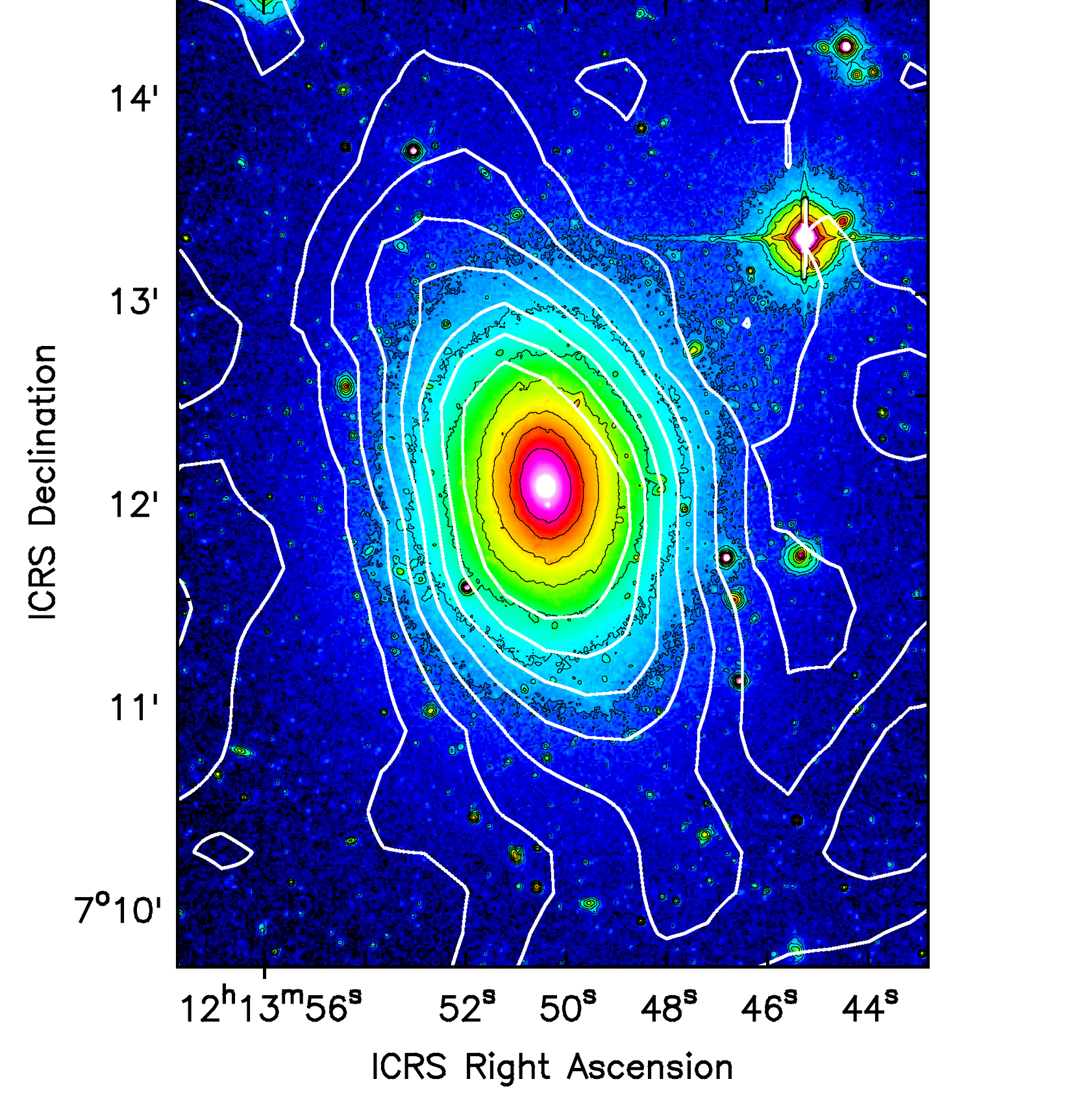}
}
\caption{\hi\ distribution in NGC~4191; the left panel shows the column density in
greyscale and contours, with contours at (0.1, 0.2, 0.4, 0.6, and 0.8) times the peak, which is 0.876
\jybkms\ or 3.6\e{20}\persqcm.  A small magenta cross marks the optical nucleus, and the white box
marks the region displayed in the right panel.
On the right, the \hi\ distribution (white contours) is overlaid on the MATLAS $g$ image (colour and black
contours, which are spaced at a factor of 2 in surface brightness).}
\label{fig:4191m0}
\end{figure*}

The \hi\ in NGC~4191 is clearly peaked on the optical galaxy nucleus, with a peak column density of
3.6\e{20}\persqcm\ at this 9 kpc spatial resolution (Figure~\ref{fig:4191m0}).  \hi\ column densities are high in a bright
elongated ridge which is approximately 3.3\arcmin\ (38 kpc) in diameter and is poorly resolved in the minor
axis direction.  This ridge is surrounded by an
extended disc of column density $\approx$ 1.2\e{20}\persqcm, 
extending to a radius of 6\arcmin\ (68 kpc) on the north side and 4\arcmin\ (45 kpc) on the south side.
The outer disc is asymmetric, having higher column densities on the south side but a larger physical
extent on the north side.

\begin{figure*}
\includegraphics[scale=0.6,trim=0cm 1cm 0cm 0cm]{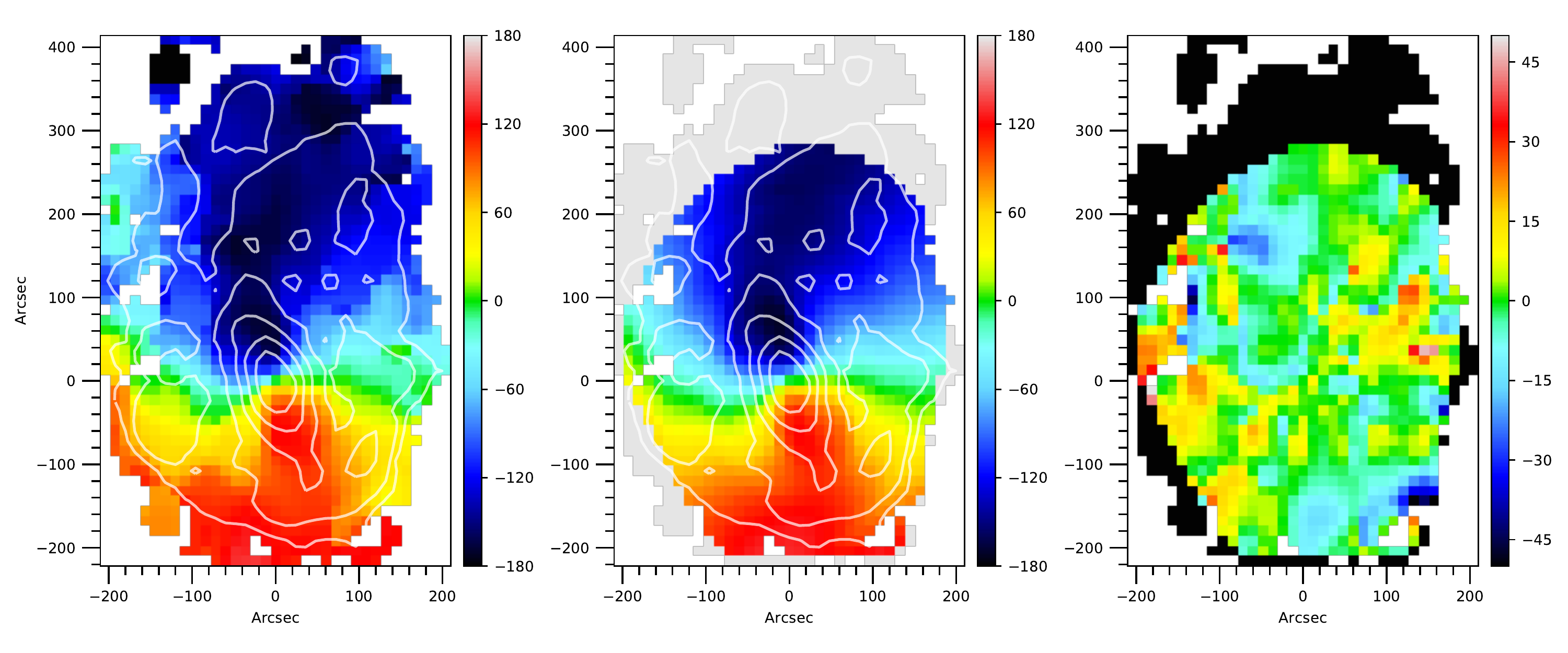}
\caption{\hi\ kinematics in NGC~4191.  Left: \hi\ velocity field, the input for the kinemetry.  The velocity field was formed by
fitting a Gaussian to the spectrum at each pixel.  Column density contours, from Figure
\ref{fig:4191m0}, are overlaid in white.  Center: reconstructed kinemetric
model velocity field.  Right: residuals.  
Each panel also has a colour wedge indicating velocities in \kms.}
\label{fig:4191modelvels}
\end{figure*}

\begin{figure}
\includegraphics[scale=0.6,trim=4cm 1cm 0cm 1cm]{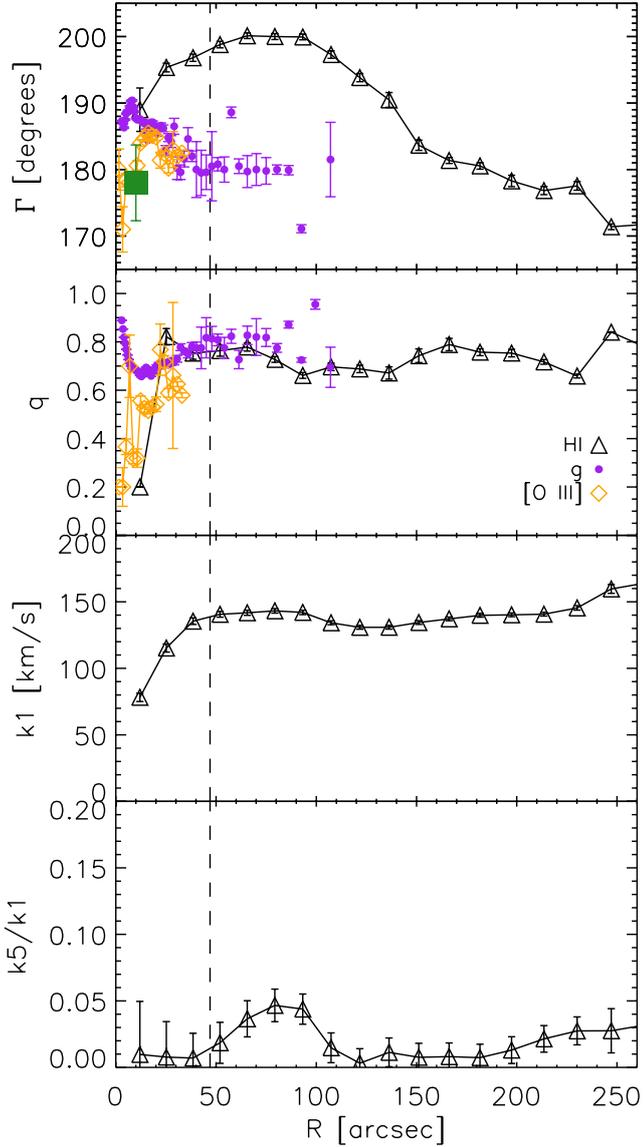}
\caption{Kinemetry analysis on the \hi\ velocity field in NGC~4191 (black triangles).  The parameter
$\Gamma$ is the position angle of the major axis, and $q$ is the ellipse axis ratio.
In the case of a thin, inclined disc with pure circular rotation, the parameter $k1$ is equivalent to the 
projected velocity $V_c \sin i$; $k5$ indicates deviations from simple circular rotation.
The dashed vertical line indicates the HI
resolution (FWHM), so values interior to this radius will be severely affected by beam smearing.  
Purple symbols indicate the stellar photometric position angle and axis ratio, from Figure
\ref{fig:4191photfitpars}; orange symbols show [O III] kinematics; and the green symbol indicates the kinematic position angle of the ionized
gas in the SAURON field of view \citep{davis11}.}
\label{fig:4191fitpars}
\end{figure}

We employ the kinemetric analysis of \citet{kinemetry} to describe the \hi\ velocity field, and the
results are in
Figures~\ref{fig:4191modelvels} and
\ref{fig:4191fitpars}.  
Figure \ref{fig:4191modelvels} presents the observed velocity field, the modeled field and the
residuals.  Figure \ref{fig:4191fitpars} presents the radial variation in the kinematic parameters of
the disc -- its orientation, inclination, circular velocity and deviations from circular velocity.
We find that at all radii unaffected by beam smearing (outside the FWHM),
the projected circular velocity $k_1$ is approximately constant at 140 \kms\ $\pm$ 5 \kms.  Here the error estimate
describes the dispersion about the mean, rather than the uncertainty in the mean. 
The inclination of the \hi\ disc can be estimated from the axis
ratio parameter $q$, for which we find a median value of 0.72 and a dispersion of 0.05.
Assuming a thin disc, the inclination of the \hi\ is 44\arcdeg\ $\pm$
4\arcdeg\ and the deprojected circular velocity is then $202^{+18}_{-14}$ \kms.  
The most prominent kinematic feature in the \hi\ disc is 
a twist in kinematic PA, which changes from 200\arcdeg\ at radii of 60\arcsec\ to 90\arcsec\ (11 to 17 kpc) to
180\arcdeg\ at radii $>$ 150\arcsec\ (28 kpc).  
Fitting the full three-dimensional \hi\ data cube with the 3DBarolo software \citep{barolo} reproduces
all these results, with a slightly larger inclination 49 \arcdeg\ $\pm$ 4\arcdeg\ and correspondingly
lower rotation velocity $184 \pm 12$ \kms.
The inner kinematic PA of 200\arcdeg\ matches the photometric major axis 
of the \hi\ ridge to within a few degrees and the location of the kinematic twist 
matches the end of the ridge.  This twist is apparently not
associated with a significant change in the inclination.

\begin{figure}
\includegraphics[scale=0.6,trim=3.5cm 1cm 0cm 1cm, clip]{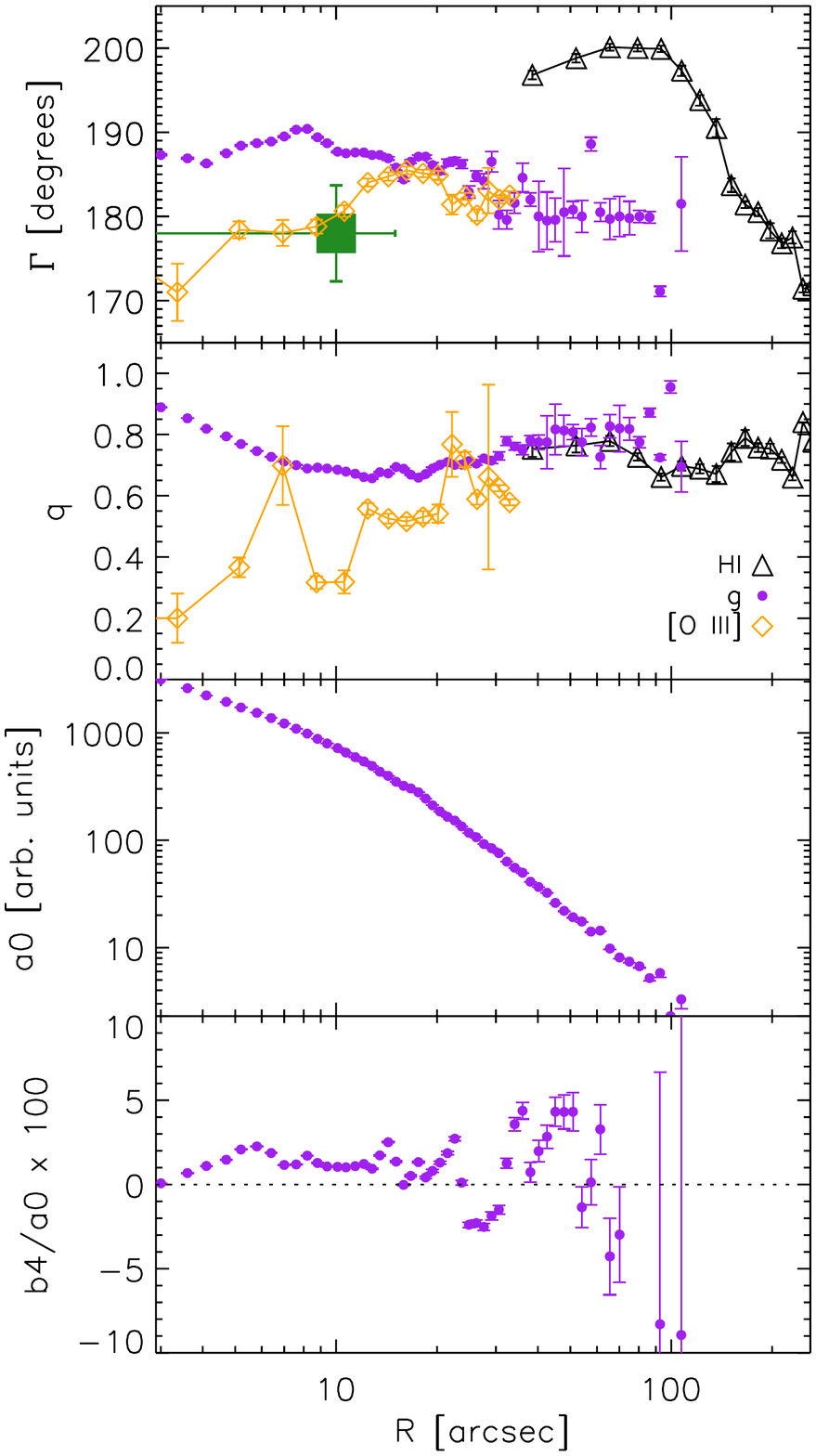}
\caption{Photometric analysis on the deep MATLAS $g$ image of NGC~4191.  Symbols are as in Figure
\ref{fig:4191fitpars}, and $\Gamma$ and $q$ are again the ellipse major axis position angle and
axis ratio.  The parameter $a0$ is the stellar surface brightness and $b4/a0$ indicates the amplitude
of the $\cos 4\theta$ component in a Fourier decomposition of the surface brightness.  Boxy isophotes
have $b4 < 0$ and discy isophotes have $b4 > 0$ \citep{jedrzejewski}.
The ionized gas kinematics for radii less than about 7\arcsec\ may be biased, due to a large mask for a
foreground star.}
\label{fig:4191photfitpars}
\end{figure}

\subsection{Stellar structure and ionized gas kinematics}

For comparison to the stellar structure of the galaxy, we have also applied the kinemetry routines to
a deep $g$-band image of NGC~4191.  The image was obtained as part of the MATLAS project \citep{duc15}.
Figure~\ref{fig:4191photfitpars} shows the fitted stellar isophote parameters as a function of the semi-major
axis length.  The galaxy has a very regular structure with a slight position angle twist from 10\arcdeg\
at a radius of 8\arcsec\ to 0\arcdeg\ at 90\arcsec, and the stellar kinematic axis is consistent with
the photometric axis \citep[at least in the inner parts, where both can be measured;][]{davor}.  
The isophote axis ratio has a minimum of 0.67 at radii between 10\arcsec\ and 20\arcsec, increasing to
0.80 outside of that range.
These values are entirely consistent with those found by \citet{coccato} from SDSS data, though we are
able to trace the structure nearly a factor of 2 farther in radius.
We also find mildly discy isophotes $(b_4 > 0)$, which, together with the axis ratio dip, is 
undoubtedly due to the influence of the stellar disc that is described by \citet{coccato} as the 
secondary or inner stellar disc.

The stellar isophotal axis ratio of NGC~4191 is relatively large, especially in comparison to other 
$2\sigma$ galaxies.  
If NGC~4191 is an
oblate fast rotator, as is argued elsewhere for the $2\sigma$ galaxies, and if we adopt the 
intrinsic axis ratio 0.25 $\pm$ 0.14 for fast rotators \citep{weijmans_shapes}, 
then the {\it minimum} apparent stellar axis ratio of 0.67 implies an inclination angle of 
50$^{+4}_{-2}$ degrees.

A kinemetric analysis of the ionized gas velocity field from \citet{coccato} is also shown in
Figure~\ref{fig:4191photfitpars}.   The ionized gas kinematic PA
is a close match to the stellar photometric PA at radii $\sim$ 12\arcsec\ to 30\arcsec.  The gas's kinematic axis ratio $q$ is smaller than the
stellar photometric axis ratio, as one would expect when the gas disc is thinner than the stellar body.
The annuli for radii 7\arcsec\ to 30\arcsec\ give the ionized gas axis ratio to be $q =
0.56 \pm 0.13,$ which corresponds to an inclination of 56$^{+8}_{-10}$ degrees.  The inclination of the
ionized gas disc is thus consistent with suggestions from the stellar isophote axis ratios.

\subsection{\hi\ disc vs.\ stellar discs}

The \hi\ kinematic major axis is 20\arcdeg\ offset from the stellar photometric major
axis over the radius range 40\arcsec\ -- 90\arcsec, where we have good measurements in both bands
and the \hi\ velocity field is still very regular.
Thus, the misaligned \hi\ kinematic axis suggests that the \hi\ at radii $\lesssim$ 150\arcsec\ (28.5 kpc)
may be precessing.  The orbital timescale at that radius is 870 Myr, and it might be tempting to
associate that timescale with a proposed merger or accretion event (see Section \ref{sec:4191disc}
below), but at present we cannot make a specific attribution with any known event in the galaxy's
history.
In terms of the broader context, though, a 20\arcdeg\ misalignment is mild compared to many misalignments and
twists noted in \citetalias{serra14} and \citet{morganti06}, and it would not have been notable without such high quality
optical data and relaxed outer isophotes.

Another point of interest is that the inclination of the \hi\ disc, which we infer to be 44\arcdeg\ $\pm$
4\arcdeg\ or 49\arcdeg\ $\pm$ 4\arcdeg, is consistent with values inferred from ionized gas kinematics
and stellar isophotes.  Dust features at radii out to $\sim$ 20\arcsec\ in an HST F814W image
are also broadly consistent with this lower inclination.  But it is
very different from the 
90\arcdeg\ inferred by \citet{cap:a3dJAM} from Jeans anisotropic modeling of the stellar kinematics. 
Thus the gaseous disc of NGC~4191 seems to be kinematically quite distinct from the 
stellar structures at radii $\lesssim$ 30\arcsec.
Higher resolution \hi\ data could help to show whether the \hi\ disc is strongly warped and twisted on
scales too small to resolve in the current data.

For context, again, we note that
the evidence for a moderate inclination in NGC~4191 makes it different from most of the other
$2\sigma$ galaxies in the \atlas\ sample.
The other $2\sigma$ galaxies all have JAM inclinations greater than 57\arcdeg, which is not surprising as 
it should be much easier to detect the double-disc kinematic signature when the discs are closer to
edge-on \citep{davor}.
But the other $2\sigma$ galaxies do not have large cold gas discs, so in those cases we lack the
additional information provided by the \hi\ here.
In fact, besides NGC~4191, there are 16 other \atlas\ galaxies for which we can compare JAM stellar
kinematic inclinations to the kinematic data from large \hi\ discs \citep{denheijer}.
Five of the 17 have \hi\ inclinations
that are more than 40\arcdeg\ different from their stellar JAM inclinations.\footnote{NGC~3522, NGC~4203, NGC~4262, NGC~4278, and NGC~4191.} 
Some of these differences may reflect uncertainties in the stellar kinematic modeling as their JAM
fits are described as poor quality, but others may reflect real warps and/or misaligned dark matter
halos.

\subsection{Dynamics at large radii}

Aside from the inclination issue, though, the \hi\ dynamics in NGC~4191 are consistent with the stellar dynamics.  
The deprojected \hi\ circular velocity, $202^{+18}_{-14}$ \kms, compares favorably to 
the Tully-Fisher relation in \citet{denheijer} and to other
dynamical relations found for early-type galaxies \citep[][hereafter S16]{serra_vcirc}.
Given the measured stellar dispersion $\sigma_e$ = 124 \kms\ $\pm$ 5 \kms\ \citep{cap:a3dJAM}, the \hi\ circular
velocity is 1.63 $\pm$ 0.12 times $\sigma_e$.  This value is higher than the typical ratio 1.33
$\pm$ 0.16
found by \citetalias{serra_vcirc} but within the combined uncertainties.
The \hi\ circular velocity is also 0.87 $\pm$ 0.06 times the maximum circular velocity inferred from
the stellar dynamics, which is 232 \kms\ at a radius of 2.4\arcsec; again, the \hi\ velocity is on the
high side of the typical ratio 0.73 $\pm$ 0.13 but is consistent with the dispersion in measured
values.
And because the \hi\ rotation speed is approximately constant out to such a large radius,
the average logarithmic slope of
the total density profile in NGC~4191 is 2.06 $\pm$ 0.03, close to isothermal, as is typical for the galaxies in
\citetalias{serra_vcirc}.  

Finally, it is also worth noting that the \hi\ disc in NGC~4191 is unusually large for early-type
galaxies.  We measure the circular velocity out to 250\arcsec\ or 47.5 kpc, and for 
$R_e \approx 15$\arcsec\ \citep{cap:a3dJAM,davor13a} 
that is 16 $R_e$.  The \hi\ extends even farther, out to 68 kpc, on the North side.
Of the 16 \hi-rich early-type galaxies studied by \citet{denheijer} and \citetalias{serra_vcirc}, 
none has an \hi\ velocity measured at a radius beyond 28.2 kpc, so NGC~4191's disc is at least 50\%
larger than its nearest competitor in that sense.
There are also only two other
examples with measured \hi\ rotation velocities at $\sim 15\;R_e$.  These are NGC~2685 and NGC~3838,
and the \hi\ distribution in
NGC~3838 is noticeably less relaxed and regular than that of NGC~4191.

The total mass implied by the
\hi\ circular velocity in NGC~4191 (202 \kms\ at 47.5 kpc) is 4.5\e{11} \solmass, a factor of 9 larger than the
dynamical mass implied by the JAM analysis of \citet{cap:a3dJAM}.  The JAM dynamical masses refer to a
region of radius $\sim R_e,$ so the difference clearly implies large dark
matter contents in the outer halo of the galaxy.  And again, because of the large radius of the \hi\
disc,
this factor of 9 is unusual for \hi-rich early-type galaxies.  The data in \citetalias{serra_vcirc} show two other galaxies
(NGC~2685 and NGC~5582) whose \hi-derived dynamical masses are a factor of 6 larger than their
stellar-derived JAM masses.  NGC~4191 is even more strongly dark matter-dominated than these.  For
comparison, the median value in \citetalias{serra_vcirc} is a factor of 2.2 and the dispersion is 
similarly a factor of 2.1.
Thus, NGC~4191 should provide an interesting case study for dynamical modeling at large radii in an
early-type galaxy.

\subsection{Discussion: Formation and evolution of NGC~4191 and other $2\sigma$
galaxies}\label{sec:4191disc}

\citet{davor13a} described the
surface brightness profile of NGC~4191 as fit by a S\'ersic profile of index 3.2 $\pm$ 0.2.
\citet{coccato} investigated its structure in more detail and  described it as consisting of three components:
a bulge, which dominates the
surface brightness profile and the stellar velocity field at radii $< 3$\arcsec; a large disc, which
dominates the surface brightness at radii $> 22$\arcsec; and a smaller secondary disc, which
dominates the surface brightness and the velocity field at 
intermediate radii 3\arcsec $< r <$ 22\arcsec.  The secondary disc is counterrotating with respect
to the other two components, and there is some evidence that its stars are younger.
It is estimated to contain one third of the total galaxy luminosity, but since it has
younger stars, it has a somewhat smaller percentage ($\sim$ 20\%) of the total stellar mass.
The \hi\ disc rotates in the same sense as the ionized gas and the secondary stellar component.
With these results in mind, we consider models for the formation of galaxies like NGC~4191.

Two general scenarios proposed for the formation of $2\sigma$ galaxies involve 
(1) accretion and (2) mergers  
\citep[e.g.][]{crocker4550,corsini,coccato,mitzkus5102}.
Option 1 involves
the accretion of retrograde gas (and possibly stars) onto a disc galaxy.  Misaligned but roughly
retrograde gas should settle into the equatorial plane, and subsequent star formation in that
gas would form a coplanar but retrograde stellar disc. 
\citet{algorry} describe a variant in which a galaxy at the intersection of two cosmic filaments 
could accrete gas first with one spin, from one dominant filament, and then with the opposite spin 
(from another filament).
Option 2 involves a precisely aligned major merger in which two
disc galaxies and their mutual orbit are all roughly coplanar, and at least one of the galaxies' spins
is antiparallel to the orbital angular momentum. 
Just how closely aligned the angular momentum vectors would have to be, in option 2, has been partially
explored but not well constrained \citep{bournaud05, bois}.

In the case of NGC~4191, \citet{coccato} support the
first option on the grounds that
they claim tentative evidence for a radial age gradient in NGC~4191, and they propose that a major
merger (as in option 2) would erase such gradients.  It would probably be worthwhile to check the
question more quantitatively since \citet{bois} find that the discy structure of the larger progenitor
can be preserved mostly intact in an aligned merger of this type.
In any case, the large \hi\ disc in NGC~4191 certainly could be evidence of the
accreted retrograde gas.  It is worth noting that the present gas mass, 5\e{9} \solmass\ including
helium, is
smaller than the value that \citet{coccato} have estimated for the retrograde stellar disc
($\sim 10^{10}$ \solmass).  The CO nondetection also implies a molecular mass $< 10^8$ \solmass\ 
\citep{a3dco}.  Thus if the \hi\ disc is the remnant of a disc growth phase as in option 1 above, the
original mass of accreted gas must have been several times larger than the current gas mass.  In fact the
mass of accreted gas would probably have been similar to the stellar mass of the galaxy at the time.

Taking a broader view, we note that
the configuration in NGC~4191, where the cold gas is rotating in the same sense as the
disc with the smaller scale length, is apparently the most common configuration for $2\sigma$
early-type galaxies with cold gas.
In the \atlas\ sample 11 examples of $2\sigma$ galaxies were identified \citep{davor}, and three of them have
cold gas with interferometric maps.
These are IC~0719 \citep{a3dcocarma}, NGC~4550 \citep{crocker4550}, and NGC~4191.
Beyond \atlas, NGC~5102 is also an early-type 2$\sigma$ galaxy with cold gas \citep{mitzkus5102}. 
Three of these four cases match the configuration in NGC~4191.\footnote{Additional observations of NGC~4476 from the PPAK IFU at Calar Alto show that this galaxy is also a
2$\sigma$ galaxy and the field of view of the SAURON instrument is slightly too small to show the
counterrotation.  NGC~4476 also fits the pattern where the cold and ionized gas rotate like the inner
stellar disc \citep{alisonthesis}.  If NGC~4476 is included, then four of the five cases with cold
gas maps match the configuration in NGC~4191.}
The outlier is NGC~4550, 
which is found to have two discs of similar scale lengths but different scale heights, and where the
molecular and ionized gas rotate like the thicker disc.  \citet{crocker4550} support the merger interpretation
(option 2 above) for NGC~4550 and provide a numerical simulation to justify its feasibility.
If the configuration found in the other three or four cases can be associated with the accretion scenario
(option 1 above), then these data suggest that the accretion scenario is the more common formation
pathway for 2$\sigma$ galaxies.
It could also be the case that there is an observational bias, such that $2\sigma$ galaxies are easier
to detect when the counterrotating disc is younger and brighter, as might be associated with the
accretion scenario.

The $2\sigma$ galaxies may also be extreme examples of the general phenomenon of gas accretion and
disc growth in early-type galaxies.  The SAURON and \atlas\ surveys contain several other 
cases of fast rotators that host compact kinematically decoupled cores (KDCs), and in some of these
cases there is also good evidence that the KDCs may have
grown out of accreted gas.  Here we consider NGC~3032 and NGC~4150 \citep{mcdermid_oasis, ybc,
lucero13, oosterloo10, morganti06}.  
In NGC~3032 the sense of rotation of the molecular and atomic gas matches that of the
young stellar KDC, and it is retrograde with respect to the large-scale stellar disc.
In NGC~4150 the cold gas rotates in the same sense as the large-scale stellar disc, and it is
therefore the opposite sense to that of the young stellar KDC; this case is confusing and still poorly
understood, and it may involve two separate episodes of accretion.  However NGC~3032 might be
an example of an early-stage modest $2\sigma$ galaxy, where most of the accreted, retrograde gas is still in
molecular and atomic form.  There are several other cases in \atlas\ with retrograde gas
\citep{davis_pv}, but
the quantity of cold gas is usually not large enough to make a stellar disc with more
than a few percent of the total stellar mass.  Thus, as discussed above for NGC~4191 specifically,
if the $2\sigma$ galaxies represent cases of disc growth from accreted retrograde gas, they
would require extreme
accretion events much more massive than the recent ones.  The scenario could potentially be tested in
cosmological simulations with large volumes and accurate treatment of gas cooling.

\section{SUMMARY}\label{sec:summary}

We present new VLA \hi\ observations of 5 \atlas\ slow rotator early-type galaxies, plus newly reduced
archival data on 4 others.
Three other \atlas\ fast rotators are also present in the observed fields along with numerous spirals
and dwarf galaxies, and we find one \hi\ cloud without an apparent optical counterpart.
With these new data, 34 of the 36 \atlas\ slow rotators now have
sensitive \hi\ observations at arcminute-scale resolution (4 to 10 kpc)
from either WSRT, the VLA, or Arecibo.
Typical column density limits are about 2.5\e{19}\persqcm, and typical mass limits are a few $10^7$
\solmass\ of HI. 

The \atlas\ survey is the first large, volume-limited survey of stellar kinematics in early-type
galaxies.  Thus it provides the best opportunity to make the connections between the galaxies' histories,
as recorded in their stellar kinematics, with their histories as recorded in their cold gas. 
The new observations also add to what is known about the diversity of \hi\ in slow rotators, as the two
new detections are unsual for slow rotators.

The new data enable a more complete assessment of the \hi\ properties of slow and fast rotators; in
general, they reinforce the similarities between the \hi\ of the two types.
The \hi\ detection rates are entirely consistent with each other.
All of the \hi\ morphological types (small discs, large discs, isolated clouds and unsettled
distributions) are now known to be present in both slow and fast rotators.
The two new \hi\ detections discussed here also have very large \hi\ contents, giving M(\hi)/M$_\star \sim
0.1,$ which is nearly as high as the most gas-rich fast rotators.
As expected, these \hi-rich slow rotators are found in low density environments.
And since the slow and fast rotators probably had very different formation paths
\citep{cappellari_araa}, the fact that their \hi\ properties are very similar suggests that their current
\hi\ properties are more a reflection of their recent accretion/interaction history than their
high- and moderate-redshift formation events.

One interesting deviation from the previous picture is that the new \hi\ detections have their HI
kinematics well aligned with their stellar kinematics.  Earlier detections of relaxed \hi\ discs in
slow rotators were in roughly polar configurations, with the \hi\ kinematic axis 90\arcdeg\ out from the
stellar kinematic axis.  \citetalias{serra14} commented that it was difficult for their simulations to produce so many
polar gas discs.  The new detections alleviate that difficulty to some extent, though they do not
eliminate it.  In these senses the new \hi\ data broaden the known diversity of \hi\ properties in slow
rotators, and it would be useful to re-examine the \hi\ properties of simulated slow rotators.

NGC~1222, one of the two slow rotators that we discuss in some detail, is a starbursting merger
remnant.
Its stellar body shows prolate-like rotation and the \hi\ occurs in a moderate-inclination disc, with its
kinematic axis roughly aligned with the stars and the ionized gas.  If the galaxy is indeed prolate,
the \hi\ is in its equatorial plane.
NGC~1222 also shows shells and tidal tails, evidence of the recent merger; no \hi\ is found associated
with the shells or tails, so deeper \hi\ observations would be necessary to tell whether the merger was
gas-rich or whether perhaps the gas was accreted later.  
The question is interesting in the light of simulation work probing the origins of prolate-like
rotation and prolate shapes \citep{li2017,tsatsi2017,ebrova}.

The other slow rotator discussed in detail is NGC~4191, a ``2 $\sigma$-peak" galaxy made of two
coplanar but counterrotating stellar discs.  We find that it has an unusually 
large \hi\ disc of 
diameter greater than 100 kpc, which allows us to measure a dynamical mass out to 16 $R_e$. 
The \hi\ disc shows a mild kinematic twist but there is a large discrepancy between the $\sim$ 45\arcdeg\
inclination inferred from \hi\ and ionized gas and the $\sim$ 90\arcdeg\ inclination inferred from
an axisymmetric stellar dynamical model.  Inclinations aside, the \hi\ disc rotates in the same sense as
the smaller and more compact of the two stellar discs, so it is retrograde with respect to the primary
stellar disc.  This configuration is found in three of the four 2$\sigma$ galaxies with the appropriate
data.
Accretion-driven disc growth might explain the retrograde gas and stellar disc in NGC~4191, though more detailed
simulations of the aligned merger model should also be carried out.
There is abundant evidence for misaligned cold gas accretion in the \atlas\ sample, but NGC~4191 has
the largest cold gas mass of these cases.

In any discussion of slow rotators it is useful to keep in mind that there is some argument for
removing the 2$\sigma$ galaxies like NGC~4191 from the slow rotator class
\citep[e.g.][]{cappellari_araa}.  We have retained the 2$\sigma$ galaxies in the discussion 
here simply because of their past classifications.

In general, because the early-type galaxies are relatively poor in cold gas, their cold gas properties
provide sensitive benchmarks for numerical simulations of galaxy evolution.  Their diversity in gas
content and in gas kinematics stands in striking contrast to the properties of spirals, and therefore
the early-type galaxies provide a greater challenge to the simulations that include cold gas
\citep[e.g.][]{serra14,lagos15}.

\section*{Acknowledgments}

Thanks to L.\ Coccato for providing the \oiii\ data from their paper, and to Alison Crocker and
Gareth C.\ Jones for helpful discussions.

The National Radio Astronomy Observatory is a facility of the National Science Foundation operated under cooperative agreement by Associated Universities, Inc.
This research has made use of the NASA/IPAC Extragalactic Database (NED) which is operated by the Jet Propulsion Laboratory, California Institute of Technology, under contract with the National Aeronautics and Space Administration. 
We acknowledge the usage of the HyperLeda database (http://leda.univ-lyon1.fr).

\bibliographystyle{mnras}
\bibliography{myrefs}

\begin{thebibliography}{}
\makeatletter
\relax
\def\mn@urlcharsother{\let\do\@makeother \do\$\do\&\do\#\do\^\do\_\do\%\do\~}
\def\mn@doi{\begingroup\mn@urlcharsother \@ifnextchar [ {\mn@doi@}
  {\mn@doi@[]}}
\def\mn@doi@[#1]#2{\def\@tempa{#1}\ifx\@tempa\@empty \href
  {http://dx.doi.org/#2} {doi:#2}\else \href {http://dx.doi.org/#2} {#1}\fi
  \endgroup}
\def\mn@eprint#1#2{\mn@eprint@#1:#2::\@nil}
\def\mn@eprint@arXiv#1{\href {http://arxiv.org/abs/#1} {{\tt arXiv:#1}}}
\def\mn@eprint@dblp#1{\href {http://dblp.uni-trier.de/rec/bibtex/#1.xml}
  {dblp:#1}}
\def\mn@eprint@#1:#2:#3:#4\@nil{\def\@tempa {#1}\def\@tempb {#2}\def\@tempc
  {#3}\ifx \@tempc \@empty \let \@tempc \@tempb \let \@tempb \@tempa \fi \ifx
  \@tempb \@empty \def\@tempb {arXiv}\fi \@ifundefined
  {mn@eprint@\@tempb}{\@tempb:\@tempc}{\expandafter \expandafter \csname
  mn@eprint@\@tempb\endcsname \expandafter{\@tempc}}}

\bibitem[\protect\citeauthoryear{{Alatalo} et~al.,}{{Alatalo}
  et~al.}{2013}]{a3dcocarma}
{Alatalo} K.,  et~al., 2013, \mn@doi [\mnras] {10.1093/mnras/sts299}, \href
  {http://adsabs.harvard.edu/abs/2013MNRAS.432.1796A} {432, 1796}

\bibitem[\protect\citeauthoryear{{Algorry}, {Navarro}, {Abadi}, {Sales},
  {Steinmetz}  \& {Piontek}}{{Algorry} et~al.}{2014}]{algorry}
{Algorry} D.~G.,  {Navarro} J.~F.,  {Abadi} M.~G.,  {Sales} L.~V.,  {Steinmetz}
  M.,   {Piontek} F.,  2014, \mn@doi [\mnras] {10.1093/mnras/stt2154}, \href
  {http://adsabs.harvard.edu/abs/2014MNRAS.437.3596A} {437, 3596}

\bibitem[\protect\citeauthoryear{{Bois} et~al.,}{{Bois} et~al.}{2011}]{bois}
{Bois} M.,  et~al., 2011, \mn@doi [\mnras] {10.1111/j.1365-2966.2011.19113.x},
  \href {http://adsabs.harvard.edu/abs/2011MNRAS.416.1654B} {416, 1654}

\bibitem[\protect\citeauthoryear{{Boizelle}, {Barth}, {Darling}, {Baker},
  {Buote}, {Ho}  \& {Walsh}}{{Boizelle} et~al.}{2017}]{boizelle}
{Boizelle} B.~D.,  {Barth} A.~J.,  {Darling} J.,  {Baker} A.~J.,  {Buote}
  D.~A.,  {Ho} L.~C.,   {Walsh} J.~L.,  2017, \mn@doi [\apj]
  {10.3847/1538-4357/aa8266}, \href
  {http://adsabs.harvard.edu/abs/2017ApJ...845..170B} {845, 170}

\bibitem[\protect\citeauthoryear{{Boselli}, {Cortese}  \& {Boquien}}{{Boselli}
  et~al.}{2014}]{boselli2014}
{Boselli} A.,  {Cortese} L.,   {Boquien} M.,  2014, \mn@doi [\aap]
  {10.1051/0004-6361/201322311}, \href
  {http://adsabs.harvard.edu/abs/2014A%26A...564A..65B} {564, A65}

\bibitem[\protect\citeauthoryear{{Bournaud}, {Jog}  \& {Combes}}{{Bournaud}
  et~al.}{2005}]{bournaud05}
{Bournaud} F.,  {Jog} C.~J.,   {Combes} F.,  2005, \mn@doi [\aap]
  {10.1051/0004-6361:20042036}, \href
  {http://adsabs.harvard.edu/abs/2005A%26A...437...69B} {437, 69}

\bibitem[\protect\citeauthoryear{{Brauher}, {Dale}  \& {Helou}}{{Brauher}
  et~al.}{2008}]{brauher}
{Brauher} J.~R.,  {Dale} D.~A.,   {Helou} G.,  2008, \mn@doi [\apjs]
  {10.1086/590249}, \href {http://adsabs.harvard.edu/abs/2008ApJS..178..280B}
  {178, 280}

\bibitem[\protect\citeauthoryear{{Cappellari}}{{Cappellari}}{2016}]{cappellari_araa}
{Cappellari} M.,  2016, \mn@doi [\araa] {10.1146/annurev-astro-082214-122432},
  \href {http://adsabs.harvard.edu/abs/2016ARA%26A..54..597C} {54, 597}

\bibitem[\protect\citeauthoryear{{Cappellari} et~al.,}{{Cappellari}
  et~al.}{2011a}]{cappellari_a3d1}
{Cappellari} M.,  et~al., 2011a, \mn@doi [\mnras]
  {10.1111/j.1365-2966.2010.18174.x}, \href
  {http://adsabs.harvard.edu/abs/2011MNRAS.413..813C} {413, 813}

\bibitem[\protect\citeauthoryear{{Cappellari} et~al.,}{{Cappellari}
  et~al.}{2011b}]{cap-density}
{Cappellari} M.,  et~al., 2011b, \mn@doi [\mnras]
  {10.1111/j.1365-2966.2011.18600.x}, \href
  {http://adsabs.harvard.edu/abs/2011MNRAS.416.1680C} {416, 1680}

\bibitem[\protect\citeauthoryear{{Cappellari} et~al.,}{{Cappellari}
  et~al.}{2013}]{cap:a3dJAM}
{Cappellari} M.,  et~al., 2013, \mn@doi [\mnras] {10.1093/mnras/stt562}, \href
  {http://adsabs.harvard.edu/abs/2013MNRAS.432.1709C} {432, 1709}

\bibitem[\protect\citeauthoryear{{Coccato} et~al.,}{{Coccato}
  et~al.}{2015}]{coccato}
{Coccato} L.,  et~al., 2015, \mn@doi [\aap] {10.1051/0004-6361/201526560},
  \href {http://adsabs.harvard.edu/abs/2015A%26A...581A..65C} {581, A65}

\bibitem[\protect\citeauthoryear{{Condon}, {Cotton}, {Greisen}, {Yin},
  {Perley}, {Taylor}  \& {Broderick}}{{Condon} et~al.}{1998}]{nvss}
{Condon} J.~J.,  {Cotton} W.~D.,  {Greisen} E.~W.,  {Yin} Q.~F.,  {Perley}
  R.~A.,  {Taylor} G.~B.,   {Broderick} J.~J.,  1998, \mn@doi [\aj]
  {10.1086/300337}, \href {http://adsabs.harvard.edu/abs/1998AJ....115.1693C}
  {115, 1693}

\bibitem[\protect\citeauthoryear{{Corsini}}{{Corsini}}{2014}]{corsini}
{Corsini} E.~M.,  2014, in {Iodice} E.,  {Corsini} E.~M.,  eds,  Astronomical
  Society of the Pacific Conference Series Vol. 486, Multi-Spin Galaxies. p.~51
  (\mn@eprint {arXiv} {1403.1263})

\bibitem[\protect\citeauthoryear{{Cortese} et~al.,}{{Cortese}
  et~al.}{2016}]{cortese2016}
{Cortese} L.,  et~al., 2016, \mn@doi [\mnras] {10.1093/mnras/stw801}, \href
  {http://adsabs.harvard.edu/abs/2016MNRAS.459.3574C} {459, 3574}

\bibitem[\protect\citeauthoryear{{Courtois} \& {Tully}}{{Courtois} \&
  {Tully}}{2015}]{courtois&tully}
{Courtois} H.~M.,  {Tully} R.~B.,  2015, \mn@doi [\mnras]
  {10.1093/mnras/stu2405}, \href
  {http://adsabs.harvard.edu/abs/2015MNRAS.447.1531C} {447, 1531}

\bibitem[\protect\citeauthoryear{{Crocker}}{{Crocker}}{2009}]{alisonthesis}
{Crocker} A.~F.,  2009, PhD thesis, University of Oxford

\bibitem[\protect\citeauthoryear{{Crocker}, {Jeong}, {Komugi}, {Combes},
  {Bureau}, {Young}  \& {Yi}}{{Crocker} et~al.}{2009}]{crocker4550}
{Crocker} A.~F.,  {Jeong} H.,  {Komugi} S.,  {Combes} F.,  {Bureau} M.,
  {Young} L.~M.,   {Yi} S.,  2009, \mn@doi [\mnras]
  {10.1111/j.1365-2966.2008.14295.x}, \href
  {http://adsabs.harvard.edu/abs/2009MNRAS.393.1255C} {393, 1255}

\bibitem[\protect\citeauthoryear{{Davis} et~al.,}{{Davis}
  et~al.}{2011}]{davis11}
{Davis} T.~A.,  et~al., 2011, \mn@doi [\mnras]
  {10.1111/j.1365-2966.2011.19355.x}, \href
  {http://adsabs.harvard.edu/abs/2011MNRAS.417..882D} {417, 882}

\bibitem[\protect\citeauthoryear{{Davis} et~al.,}{{Davis}
  et~al.}{2013}]{davis_pv}
{Davis} T.~A.,  et~al., 2013, \mn@doi [\mnras] {10.1093/mnras/sts353}, \href
  {http://adsabs.harvard.edu/abs/2013MNRAS.429..534D} {429, 534}

\bibitem[\protect\citeauthoryear{{Davis} et~al.,}{{Davis}
  et~al.}{2014}]{davis14}
{Davis} T.~A.,  et~al., 2014, \mn@doi [\mnras] {10.1093/mnras/stu570}, \href
  {http://adsabs.harvard.edu/abs/2014MNRAS.444.3427D} {444, 3427}

\bibitem[\protect\citeauthoryear{{Di Teodoro} \& {Fraternali}}{{Di Teodoro} \&
  {Fraternali}}{2015}]{barolo}
{Di Teodoro} E.~M.,  {Fraternali} F.,  2015, \mn@doi [\mnras]
  {10.1093/mnras/stv1213}, \href
  {http://adsabs.harvard.edu/abs/2015MNRAS.451.3021D} {451, 3021}

\bibitem[\protect\citeauthoryear{{Doyle} et~al.,}{{Doyle}
  et~al.}{2005}]{doyle05}
{Doyle} M.~T.,  et~al., 2005, \mn@doi [\mnras]
  {10.1111/j.1365-2966.2005.09159.x}, \href
  {http://adsabs.harvard.edu/abs/2005MNRAS.361...34D} {361, 34}

\bibitem[\protect\citeauthoryear{{Duc} et~al.,}{{Duc} et~al.}{2011}]{p9-5557}
{Duc} P.-A.,  et~al., 2011, \mn@doi [\mnras]
  {10.1111/j.1365-2966.2011.19137.x}, \href
  {http://adsabs.harvard.edu/abs/2011MNRAS.417..863D} {417, 863}

\bibitem[\protect\citeauthoryear{{Duc} et~al.,}{{Duc} et~al.}{2015}]{duc15}
{Duc} P.-A.,  et~al., 2015, \mn@doi [\mnras] {10.1093/mnras/stu2019}, \href
  {http://adsabs.harvard.edu/abs/2015MNRAS.446..120D} {446, 120}

\bibitem[\protect\citeauthoryear{{Duprie} \& {Schneider}}{{Duprie} \&
  {Schneider}}{1996}]{duprie96}
{Duprie} K.,  {Schneider} S.~E.,  1996, \mn@doi [\aj] {10.1086/118067}, \href
  {http://adsabs.harvard.edu/abs/1996AJ....112..937D} {112, 937}

\bibitem[\protect\citeauthoryear{{Dwarakanath}, {van Gorkom}  \&
  {Owen}}{{Dwarakanath} et~al.}{1994}]{dwarakanath}
{Dwarakanath} K.~S.,  {van Gorkom} J.~H.,   {Owen} F.~N.,  1994, \mn@doi [\apj]
  {10.1086/174586}, \href {http://adsabs.harvard.edu/abs/1994ApJ...432..469D}
  {432, 469}

\bibitem[\protect\citeauthoryear{{Ebrova} \& {Lokas}}{{Ebrova} \&
  {Lokas}}{2017}]{ebrova}
{Ebrova} I.,  {Lokas} E.~L.,  2017, preprint, \href
  {http://adsabs.harvard.edu/abs/2017arXiv170803311E} {} (\mn@eprint {arXiv}
  {1708.03311})

\bibitem[\protect\citeauthoryear{{Emsellem} et~al.,}{{Emsellem}
  et~al.}{2011}]{emsellem_a3d}
{Emsellem} E.,  et~al., 2011, \mn@doi [\mnras]
  {10.1111/j.1365-2966.2011.18496.x}, \href
  {http://adsabs.harvard.edu/abs/2011MNRAS.414..888E} {414, 888}

\bibitem[\protect\citeauthoryear{{Giovanelli} et~al.,}{{Giovanelli}
  et~al.}{2005}]{giovanelli05}
{Giovanelli} R.,  et~al., 2005, \mn@doi [\aj] {10.1086/497431}, \href
  {http://adsabs.harvard.edu/abs/2005AJ....130.2598G} {130, 2598}

\bibitem[\protect\citeauthoryear{{Haynes} et~al.,}{{Haynes}
  et~al.}{2011}]{alfa40}
{Haynes} M.~P.,  et~al., 2011, \mn@doi [\aj] {10.1088/0004-6256/142/5/170},
  \href {http://adsabs.harvard.edu/abs/2011AJ....142..170H} {142, 170}

\bibitem[\protect\citeauthoryear{{Huchtmeier} \& {Richter}}{{Huchtmeier} \&
  {Richter}}{1989}]{hr89}
{Huchtmeier} W.~K.,  {Richter} O.-G.,  1989, {A General Catalog of HI
  Observations of Galaxies. The Reference Catalog.}

\bibitem[\protect\citeauthoryear{{Janowiecki} et~al.,}{{Janowiecki}
  et~al.}{2015}]{janowiecki}
{Janowiecki} S.,  et~al., 2015, \mn@doi [\apj] {10.1088/0004-637X/801/2/96},
  \href {http://adsabs.harvard.edu/abs/2015ApJ...801...96J} {801, 96}

\bibitem[\protect\citeauthoryear{{Jedrzejewski}}{{Jedrzejewski}}{1987}]{jedrzejewski}
{Jedrzejewski} R.~I.,  1987, \mn@doi [\mnras] {10.1093/mnras/226.4.747}, \href
  {http://adsabs.harvard.edu/abs/1987MNRAS.226..747J} {226, 747}

\bibitem[\protect\citeauthoryear{{Krajnovi{\'c}}, {Cappellari}, {de Zeeuw}  \&
  {Copin}}{{Krajnovi{\'c}} et~al.}{2006}]{kinemetry}
{Krajnovi{\'c}} D.,  {Cappellari} M.,  {de Zeeuw} P.~T.,   {Copin} Y.,  2006,
  \mn@doi [\mnras] {10.1111/j.1365-2966.2005.09902.x}, \href
  {http://adsabs.harvard.edu/abs/2006MNRAS.366..787K} {366, 787}

\bibitem[\protect\citeauthoryear{{Krajnovi{\'c}} et~al.,}{{Krajnovi{\'c}}
  et~al.}{2011}]{davor}
{Krajnovi{\'c}} D.,  et~al., 2011, \mn@doi [\mnras]
  {10.1111/j.1365-2966.2011.18560.x}, \href
  {http://adsabs.harvard.edu/abs/2011MNRAS.414.2923K} {414, 2923}

\bibitem[\protect\citeauthoryear{{Krajnovi{\'c}} et~al.,}{{Krajnovi{\'c}}
  et~al.}{2013a}]{davor13a}
{Krajnovi{\'c}} D.,  et~al., 2013a, \mn@doi [\mnras] {10.1093/mnras/sts315},
  \href {http://adsabs.harvard.edu/abs/2013MNRAS.432.1768K} {432, 1768}

\bibitem[\protect\citeauthoryear{{Krajnovi{\'c}} et~al.,}{{Krajnovi{\'c}}
  et~al.}{2013b}]{davor_cusp}
{Krajnovi{\'c}} D.,  et~al., 2013b, \mn@doi [\mnras] {10.1093/mnras/stt905},
  \href {http://adsabs.harvard.edu/abs/2013MNRAS.433.2812K} {433, 2812}

\bibitem[\protect\citeauthoryear{{Lagos}, {Padilla}, {Davis}, {Lacey}, {Baugh},
  {Gonzalez-Perez}, {Zwaan}  \& {Contreras}}{{Lagos} et~al.}{2015}]{lagos15}
{Lagos} C.~d.~P.,  {Padilla} N.~D.,  {Davis} T.~A.,  {Lacey} C.~G.,  {Baugh}
  C.~M.,  {Gonzalez-Perez} V.,  {Zwaan} M.~A.,   {Contreras} S.,  2015, \mn@doi
  [\mnras] {10.1093/mnras/stu2763}, \href
  {http://adsabs.harvard.edu/abs/2015MNRAS.448.1271L} {448, 1271}

\bibitem[\protect\citeauthoryear{{Lagos}, {Theuns}, {Stevens}, {Cortese},
  {Padilla}, {Davis}, {Contreras}  \& {Croton}}{{Lagos}
  et~al.}{2017}]{lagos2017}
{Lagos} C.~d.~P.,  {Theuns} T.,  {Stevens} A.~R.~H.,  {Cortese} L.,  {Padilla}
  N.~D.,  {Davis} T.~A.,  {Contreras} S.,   {Croton} D.,  2017, \mn@doi
  [\mnras] {10.1093/mnras/stw2610}, \href
  {http://adsabs.harvard.edu/abs/2017MNRAS.464.3850L} {464, 3850}

\bibitem[\protect\citeauthoryear{{Li}, {Mao}, {Emsellem}, {Xu}, {Springel}  \&
  {Krajnovi{\'c}}}{{Li} et~al.}{2018}]{li2017}
{Li} H.,  {Mao} S.,  {Emsellem} E.,  {Xu} D.,  {Springel} V.,   {Krajnovi{\'c}}
  D.,  2018, \mn@doi [\mnras] {10.1093/mnras/stx2374}, \href
  {http://adsabs.harvard.edu/abs/2018MNRAS.473.1489L} {473, 1489}

\bibitem[\protect\citeauthoryear{{Lim}, {Ao}  \& {Dinh-V-Trung}}{{Lim}
  et~al.}{2008}]{lim_1275}
{Lim} J.,  {Ao} Y.,   {Dinh-V-Trung} 2008, \mn@doi [\apj] {10.1086/523664},
  \href {http://adsabs.harvard.edu/abs/2008ApJ...672..252L} {672, 252}

\bibitem[\protect\citeauthoryear{{Lucero} \& {Young}}{{Lucero} \&
  {Young}}{2013}]{lucero13}
{Lucero} D.~M.,  {Young} L.~M.,  2013, \mn@doi [\aj]
  {10.1088/0004-6256/145/3/56}, \href
  {http://adsabs.harvard.edu/abs/2013AJ....145...56L} {145, 56}

\bibitem[\protect\citeauthoryear{{Lucero}, {Young}  \& {van Gorkom}}{{Lucero}
  et~al.}{2005}]{lucero4476}
{Lucero} D.~M.,  {Young} L.~M.,   {van Gorkom} J.~H.,  2005, \mn@doi [\aj]
  {10.1086/426750}, \href {http://adsabs.harvard.edu/abs/2005AJ....129..647L}
  {129, 647}

\bibitem[\protect\citeauthoryear{{McConnachie}}{{McConnachie}}{2012}]{mcconnachie}
{McConnachie} A.~W.,  2012, \mn@doi [\aj] {10.1088/0004-6256/144/1/4}, \href
  {http://adsabs.harvard.edu/abs/2012AJ....144....4M} {144, 4}

\bibitem[\protect\citeauthoryear{{McDermid} et~al.,}{{McDermid}
  et~al.}{2006}]{mcdermid_oasis}
{McDermid} R.~M.,  et~al., 2006, \mn@doi [\mnras]
  {10.1111/j.1365-2966.2006.11065.x}, \href
  {http://adsabs.harvard.edu/abs/2006MNRAS.373..906M} {373, 906}

\bibitem[\protect\citeauthoryear{{Mitzkus}, {Cappellari}  \&
  {Walcher}}{{Mitzkus} et~al.}{2017}]{mitzkus5102}
{Mitzkus} M.,  {Cappellari} M.,   {Walcher} C.~J.,  2017, \mn@doi [\mnras]
  {10.1093/mnras/stw2677}, \href
  {http://adsabs.harvard.edu/abs/2017MNRAS.464.4789M} {464, 4789}

\bibitem[\protect\citeauthoryear{{Morganti} et~al.,}{{Morganti}
  et~al.}{2006}]{morganti06}
{Morganti} R.,  et~al., 2006, \mn@doi [\mnras]
  {10.1111/j.1365-2966.2006.10681.x}, \href
  {http://adsabs.harvard.edu/abs/2006MNRAS.371..157M} {371, 157}

\bibitem[\protect\citeauthoryear{{Mundell} et~al.,}{{Mundell}
  et~al.}{2007}]{Mundell07}
{Mundell} C.~G.,  et~al., 2007, \mn@doi [\nar] {10.1016/j.newar.2006.10.002},
  \href {http://adsabs.harvard.edu/abs/2007NewAR..51...34M} {51, 34}

\bibitem[\protect\citeauthoryear{{Negri}, {Pellegrini}  \& {Ciotti}}{{Negri}
  et~al.}{2015}]{negri2015}
{Negri} A.,  {Pellegrini} S.,   {Ciotti} L.,  2015, \mn@doi [\mnras]
  {10.1093/mnras/stv968}, \href
  {http://adsabs.harvard.edu/abs/2015MNRAS.451.1212N} {451, 1212}

\bibitem[\protect\citeauthoryear{{Ogle}, {Boulanger}, {Guillard}, {Evans},
  {Antonucci}, {Appleton}, {Nesvadba}  \& {Leipski}}{{Ogle}
  et~al.}{2010}]{ogle2010}
{Ogle} P.,  {Boulanger} F.,  {Guillard} P.,  {Evans} D.~A.,  {Antonucci} R.,
  {Appleton} P.~N.,  {Nesvadba} N.,   {Leipski} C.,  2010, \mn@doi [\apj]
  {10.1088/0004-637X/724/2/1193}, \href
  {http://adsabs.harvard.edu/abs/2010ApJ...724.1193O} {724, 1193}

\bibitem[\protect\citeauthoryear{{Oosterloo} et~al.,}{{Oosterloo}
  et~al.}{2010}]{oosterloo10}
{Oosterloo} T.,  et~al., 2010, \mn@doi [\mnras]
  {10.1111/j.1365-2966.2010.17351.x}, \href
  {http://adsabs.harvard.edu/abs/2010MNRAS.409..500O} {409, 500}

\bibitem[\protect\citeauthoryear{{Russell} et~al.,}{{Russell}
  et~al.}{2017}]{russell2017}
{Russell} H.~R.,  et~al., 2017, \mn@doi [\mnras] {10.1093/mnras/stx2255}, \href
  {http://adsabs.harvard.edu/abs/2017MNRAS.472.4024R} {472, 4024}

\bibitem[\protect\citeauthoryear{{Schiminovich}, {van Gorkom}, {van der Hulst}
  \& {Kasow}}{{Schiminovich} et~al.}{1994}]{schiminovich_cena}
{Schiminovich} D.,  {van Gorkom} J.~H.,  {van der Hulst} J.~M.,   {Kasow} S.,
  1994, \mn@doi [\apjl] {10.1086/187246}, \href
  {http://adsabs.harvard.edu/abs/1994ApJ...423L.101S} {423, L101}

\bibitem[\protect\citeauthoryear{{Schiminovich}, {van Gorkom}, {van der Hulst}
  \& {Malin}}{{Schiminovich} et~al.}{1995}]{schiminovich_2865}
{Schiminovich} D.,  {van Gorkom} J.~H.,  {van der Hulst} J.~M.,   {Malin}
  D.~F.,  1995, \mn@doi [\apjl] {10.1086/187864}, \href
  {http://adsabs.harvard.edu/abs/1995ApJ...444L..77S} {444, L77}

\bibitem[\protect\citeauthoryear{{Schiminovich}, {van Gorkom}  \& {van der
  Hulst}}{{Schiminovich} et~al.}{2013}]{schiminovich2013}
{Schiminovich} D.,  {van Gorkom} J.~H.,   {van der Hulst} J.~M.,  2013, \mn@doi
  [\aj] {10.1088/0004-6256/145/2/34}, \href
  {http://adsabs.harvard.edu/abs/2013AJ....145...34S} {145, 34}

\bibitem[\protect\citeauthoryear{{Serra} et~al.,}{{Serra}
  et~al.}{2012}]{serra12}
{Serra} P.,  et~al., 2012, \mn@doi [\mnras] {10.1111/j.1365-2966.2012.20219.x},
  \href {http://adsabs.harvard.edu/abs/2012MNRAS.422.1835S} {422, 1835}

\bibitem[\protect\citeauthoryear{{Serra} et~al.,}{{Serra}
  et~al.}{2014}]{serra14}
{Serra} P.,  et~al., 2014, \mn@doi [\mnras] {10.1093/mnras/stt2496}, \href
  {http://adsabs.harvard.edu/abs/2014MNRAS.444.3388S} {444, 3388}

\bibitem[\protect\citeauthoryear{{Serra}, {Oosterloo}, {Cappellari}, {den
  Heijer}  \& {J{\'o}zsa}}{{Serra} et~al.}{2016}]{serra_vcirc}
{Serra} P.,  {Oosterloo} T.,  {Cappellari} M.,  {den Heijer} M.,   {J{\'o}zsa}
  G.~I.~G.,  2016, \mn@doi [\mnras] {10.1093/mnras/stw1010}, \href
  {http://adsabs.harvard.edu/abs/2016MNRAS.460.1382S} {460, 1382}

\bibitem[\protect\citeauthoryear{{Smith} et~al.,}{{Smith}
  et~al.}{2012}]{mwlsmith2012}
{Smith} M.~W.~L.,  et~al., 2012, \mn@doi [\apj] {10.1088/0004-637X/748/2/123},
  \href {http://adsabs.harvard.edu/abs/2012ApJ...748..123S} {748, 123}

\bibitem[\protect\citeauthoryear{{Sparke}}{{Sparke}}{1996}]{sparke_cena}
{Sparke} L.~S.,  1996, \mn@doi [\apj] {10.1086/178193}, \href
  {http://adsabs.harvard.edu/abs/1996ApJ...473..810S} {473, 810}

\bibitem[\protect\citeauthoryear{{Temi}, {Amblard}, {Gitti}, {Brighenti},
  {Gaspari}, {Mathews}  \& {David}}{{Temi} et~al.}{2017}]{temi2017}
{Temi} P.,  {Amblard} A.,  {Gitti} M.,  {Brighenti} F.,  {Gaspari} M.,
  {Mathews} W.~G.,   {David} L.,  2017, preprint, \href
  {http://adsabs.harvard.edu/abs/2017arXiv171110630T} {} (\mn@eprint {arXiv}
  {1711.10630})

\bibitem[\protect\citeauthoryear{{Thomas}, {Dunne}, {Green}, {Clemens},
  {Alexander}  \& {Eales}}{{Thomas} et~al.}{2004a}]{thomas04}
{Thomas} H.~C.,  {Dunne} L.,  {Green} D.~A.,  {Clemens} M.~S.,  {Alexander} P.,
    {Eales} S.,  2004a, \mn@doi [\mnras] {10.1111/j.1365-2966.2004.07491.x},
  \href {http://adsabs.harvard.edu/abs/2004MNRAS.348.1197T} {348, 1197}

\bibitem[\protect\citeauthoryear{{Thomas}, {Alexander}, {Clemens}, {Green},
  {Dunne}  \& {Eales}}{{Thomas} et~al.}{2004b}]{thomas04b}
{Thomas} H.~C.,  {Alexander} P.,  {Clemens} M.~S.,  {Green} D.~A.,  {Dunne} L.,
    {Eales} S.,  2004b, \mn@doi [\mnras] {10.1111/j.1365-2966.2004.07795.x},
  \href {http://adsabs.harvard.edu/abs/2004MNRAS.351..362T} {351, 362}

\bibitem[\protect\citeauthoryear{{Tsatsi}, {Lyubenova}, {van de Ven}, {Chang},
  {Aguerri}, {Falc{\'o}n-Barroso}  \& {Macci{\`o}}}{{Tsatsi}
  et~al.}{2017}]{tsatsi2017}
{Tsatsi} A.,  {Lyubenova} M.,  {van de Ven} G.,  {Chang} J.,  {Aguerri}
  J.~A.~L.,  {Falc{\'o}n-Barroso} J.,   {Macci{\`o}} A.~V.,  2017, preprint,
  \href {http://adsabs.harvard.edu/abs/2017arXiv170705130T} {} (\mn@eprint
  {arXiv} {1707.05130})

\bibitem[\protect\citeauthoryear{{Weijmans} et~al.,}{{Weijmans}
  et~al.}{2014}]{weijmans_shapes}
{Weijmans} A.-M.,  et~al., 2014, \mn@doi [\mnras] {10.1093/mnras/stu1603},
  \href {http://adsabs.harvard.edu/abs/2014MNRAS.444.3340W} {444, 3340}

\bibitem[\protect\citeauthoryear{{Werner} et~al.,}{{Werner}
  et~al.}{2014}]{werner}
{Werner} N.,  et~al., 2014, \mn@doi [\mnras] {10.1093/mnras/stu006}, \href
  {http://adsabs.harvard.edu/abs/2014MNRAS.439.2291W} {439, 2291}

\bibitem[\protect\citeauthoryear{{Wilson} et~al.,}{{Wilson}
  et~al.}{2013}]{wilson4125}
{Wilson} C.~D.,  et~al., 2013, \mn@doi [\apjl] {10.1088/2041-8205/776/2/L30},
  \href {http://adsabs.harvard.edu/abs/2013ApJ...776L..30W} {776, L30}

\bibitem[\protect\citeauthoryear{{Young}, {Bureau}  \& {Cappellari}}{{Young}
  et~al.}{2008}]{ybc}
{Young} L.~M.,  {Bureau} M.,   {Cappellari} M.,  2008, \mn@doi [\apj]
  {10.1086/529019}, \href {http://adsabs.harvard.edu/abs/2008ApJ...676..317Y}
  {676, 317}

\bibitem[\protect\citeauthoryear{{Young} et~al.,}{{Young} et~al.}{2011}]{a3dco}
{Young} L.~M.,  et~al., 2011, \mn@doi [\mnras]
  {10.1111/j.1365-2966.2011.18561.x}, \href
  {http://adsabs.harvard.edu/abs/2011MNRAS.414..940Y} {414, 940}

\bibitem[\protect\citeauthoryear{{Young} et~al.,}{{Young}
  et~al.}{2014}]{a3d_cmd}
{Young} L.~M.,  et~al., 2014, \mn@doi [\mnras] {10.1093/mnras/stt2474}, \href
  {http://adsabs.harvard.edu/abs/2014MNRAS.444.3408Y} {444, 3408}

\bibitem[\protect\citeauthoryear{{den Heijer} et~al.,}{{den Heijer}
  et~al.}{2015}]{denheijer}
{den Heijer} M.,  et~al., 2015, \mn@doi [\aap] {10.1051/0004-6361/201526879},
  \href {http://adsabs.harvard.edu/abs/2015A%26A...581A..98D} {581, A98}

\bibitem[\protect\citeauthoryear{{van de Voort}, {Davis}, {Kere{\v s}},
  {Quataert}, {Faucher-Gigu{\`e}re}  \& {Hopkins}}{{van de Voort}
  et~al.}{2015}]{freeke_misalign}
{van de Voort} F.,  {Davis} T.~A.,  {Kere{\v s}} D.,  {Quataert} E.,
  {Faucher-Gigu{\`e}re} C.-A.,   {Hopkins} P.~F.,  2015, \mn@doi [\mnras]
  {10.1093/mnras/stv1217}, \href
  {http://adsabs.harvard.edu/abs/2015MNRAS.451.3269V} {451, 3269}

\makeatother
\end{thebibliography}

\clearpage

\appendix

\section{Other \hi\ detections}

\begin{table*}
\centering
\caption{Other \hi\ detections.}
\label{tab:otherhi}
\begin{tabular}{llcccccr}
\hline
Name & Dist. & Primary Target & \hi\ flux & \hi\ mass & Offset & PB gain & Kin. PA  \\
 &     (Mpc) &                & (\jykms) & (\solmass) & (\arcmin) &      & ($\deg$) \\  
(1) & (2) & (3) & (4) & (5) & (6) & (7) & (8) \\
\hline
NGC~5806 & 21.4 & NGC~5813 &  5.9 (0.5) & 6.4\e{8} & 21 & 0.19 -- 0.33 & 174 (3)  \\ 
NGC~5811 & 20?  & NGC~5813 & 1.07 (0.06) & 1.0\e{8} & 12 & 0.63 -- 0.73 & $-85$ (30) \\
UGC~09661 & 20? & NGC~5813 & 2.4 (0.1)  & 2.3\e{8} & 15 & 0.44 -- 0.56 & 234 (10)  \\
UGC~07945 & 40? & NGC~4690 & 2.0 (0.1)  & 7.6\e{8} & 14 & 0.62 -- 0.51 & $-81$ (5)  \\
NGC~5577 & 24.5 & NGC~5576 & 11.0 (0.2) & 1.6\e{9} & 11 & 0.67 -- 0.83 & 67 (1)   \\
AGC~249460 & 24.5? & NGC~5576 & 0.9 (0.1) & 1.3\e{8} & 18 & 0.39 & \nodata   \\ 
PGC 1218738 & 24? & NGC~5838/5846 group & 0.82 (0.09) & 1.1\e{8} & \nodata & \nodata & $-16$ (11)  \\ 
PGC 1215798 & 24? & NGC~5838/5846 group & 3.91 (0.17) & 5.3\e{8} & \nodata & \nodata & 26 (3)  \\ 
PGC 1180802 & 24? & NGC~5831/5846 group & 1.44 (0.14) & 2.0\e{8} & \nodata & \nodata & 67 (12)  \\ 
PGC 054045  & 24? & NGC~5846 & 0.28 (0.07) & 3.8\e{7} & \nodata & \nodata & \nodata   \\ 
PGC 1190358 & 24? & NGC~5831/5846 group & 0.31 (0.06) & 4.2\e{7} & \nodata & \nodata & \nodata   \\ 
PGC 1211621 & 24? & NGC~5838/5846 group & 0.21 (0.03) & 2.8\e{7} & \nodata & \nodata & \nodata   \\ 
\hline
\end{tabular}

Distances are taken from \citet{cappellari_a3d1}, though those marked
with `?' are estimated by association or from a Hubble flow calculation.  
The primary target in column 3 indicates the nearest \atlas\ galaxy and the dataset in which the object
is found.
For the objects observed with a single pointing, the offset in column 6 describes the distance from the pointing center
to the object, 
and the primary beam gain over the region of interest (possibly a range, due to the
finite source size) is in column 7.  The offset information is included to give a rough indication of
the imaging quality and the uncertainties in the measured fluxes.  
It is difficult to be precise, but uncertainties of 20\% or more may be reasonable for objects beyond
the half-power point of the primary beam (radius $\sim$ 15.5\arcmin).
The kinematic position angle in column 8 is defined from North
towards East to the {\it receding}
side of the galaxy, and it is quantified from the first velocity moment image as in \citet{kinemetry}.
\end{table*}

\begin{figure*}
\includegraphics[scale=0.7,trim=0cm 0cm 0cm 0cm]{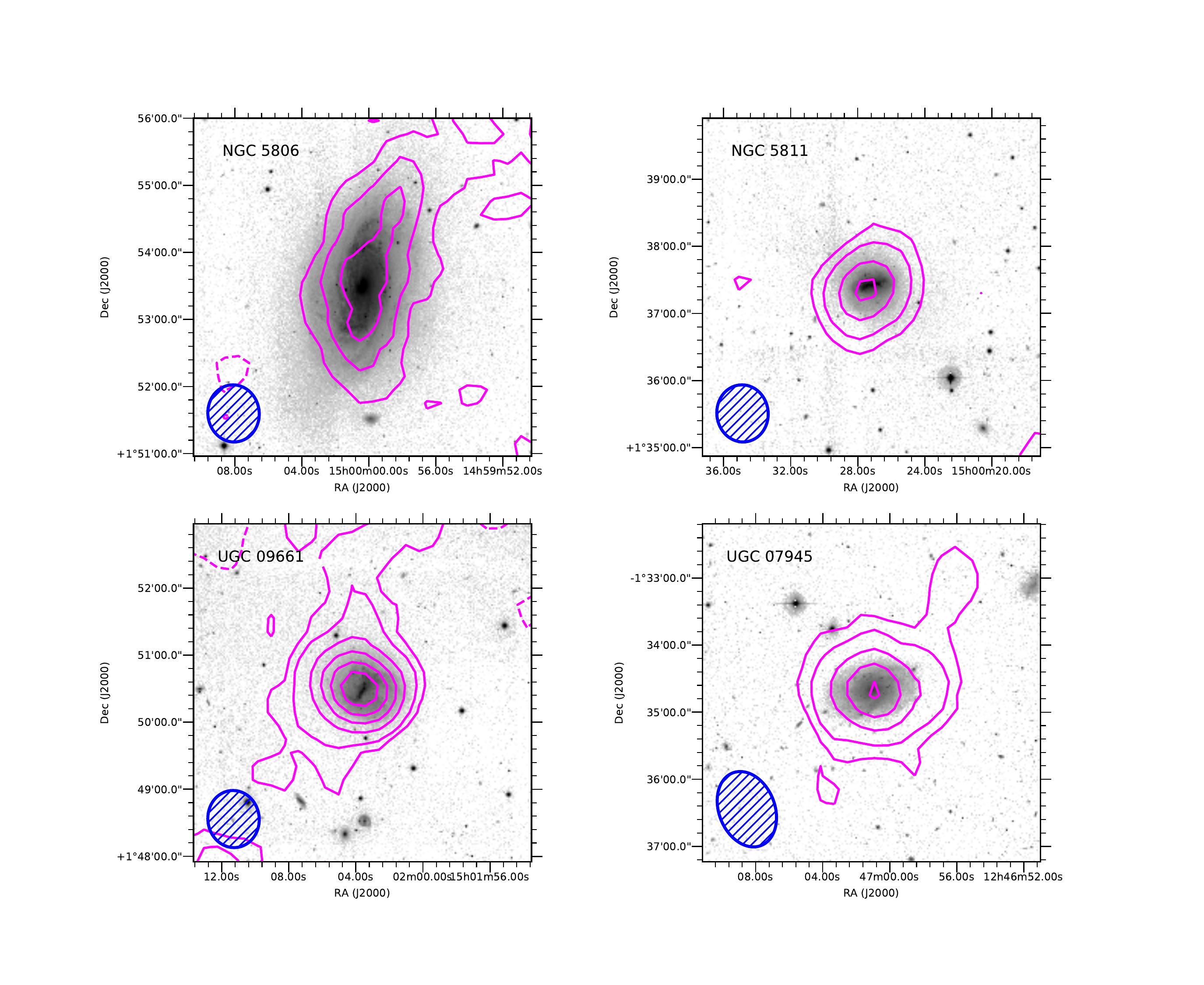}
\caption{\hi\ and optical images of other detections in the fields of NGC~5813 and NGC~4690.  
The optical images are MATLAS $g$.  The \hi\ column density contours 
are overlaid; contour levels are $(-0.5, 0.5, 1, 2, 3, 4, 5) \times 10^{20}$
\persqcm, except for NGC~5806, where they are $(-2, 2, 4, 6) \times 10^{20}$ \persqcm.
The beam size
is indicated in the bottom left corner.  Contours extending off the edges of the images are
not believed to indicate real emission, but are probably indicative of noise amplified by the primary
beam correction.}
\label{fig:others_m0}
\end{figure*}

\begin{figure*}
\includegraphics[scale=1.0]{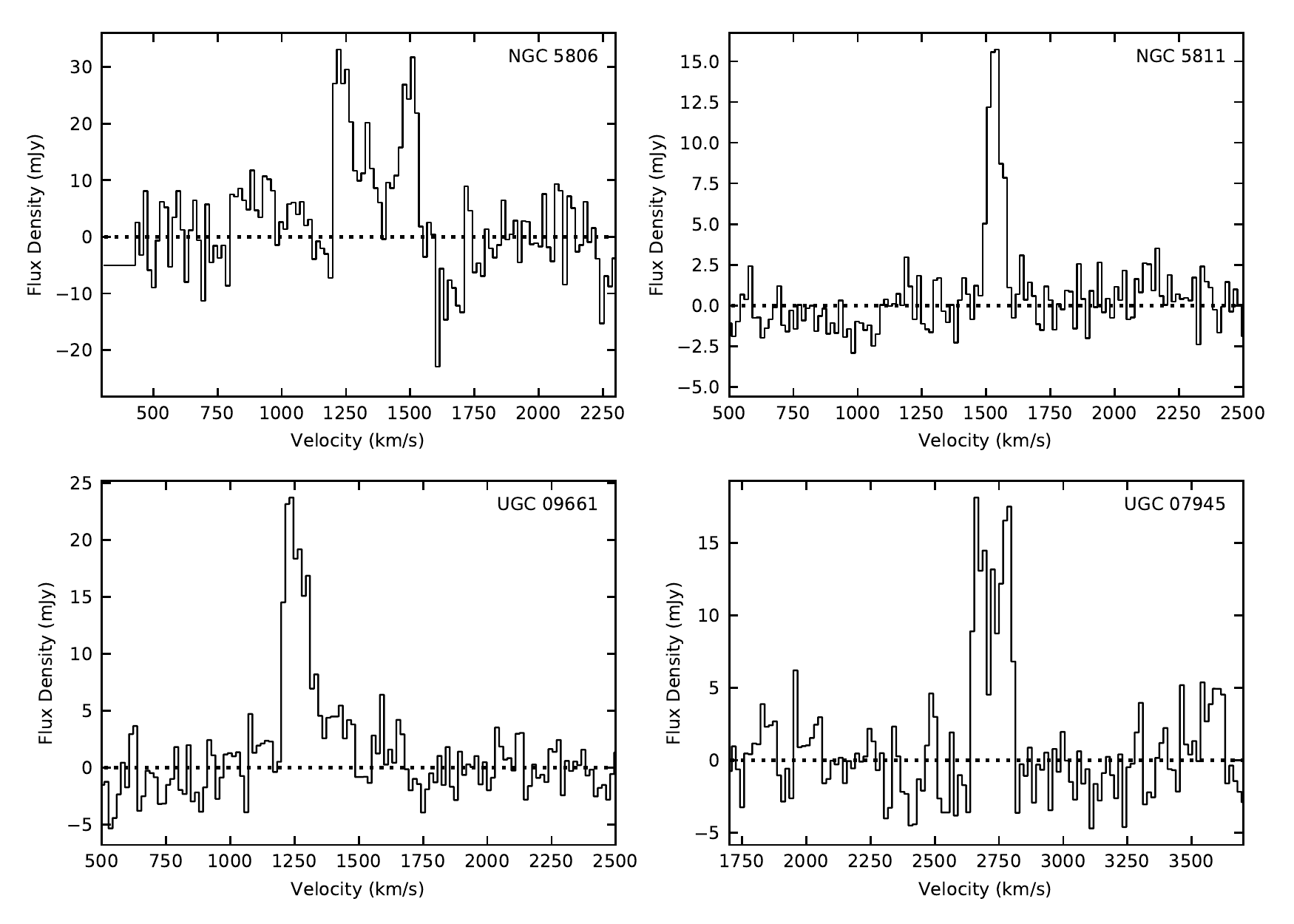}
\caption{\hi\ spectra for the galaxies in Figure~\ref{fig:others_m0}.
Spectra are constructed by using the integrated intensity to define a spatial region, and then
integrating the primary beam-corrected data cube within that same spatial region for all velocities.  
\hi\ fluxes are measured from the spectra, using the rms in the spectrum and the number of
channels with emission to define the uncertainty in the \hi\ flux.}
\label{fig:morehispec}
\end{figure*}

\begin{figure*}
\includegraphics[]{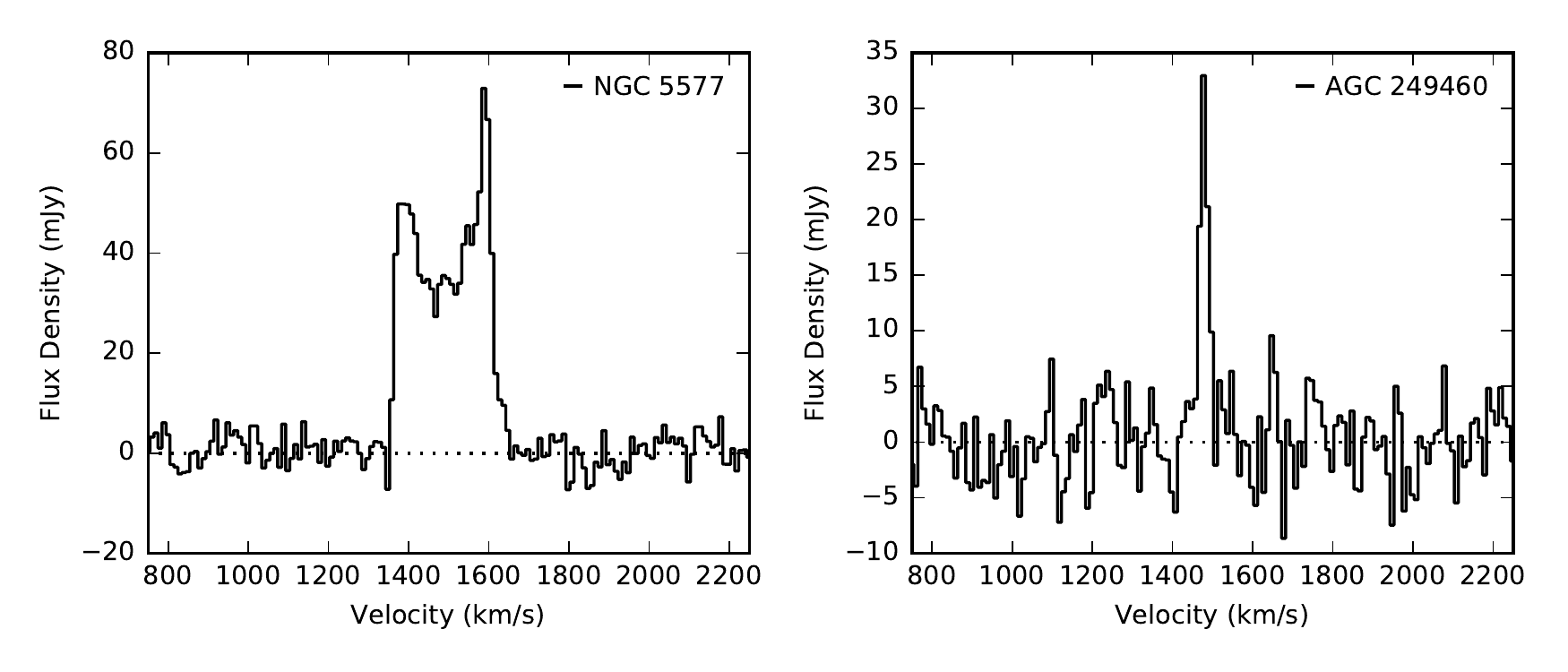}
\caption{Spectra of NGC~5577 and AGC~249460.  (The integrated intensity images are in Figure
\ref{fig:5577m0}.)}
\label{fig:5577spec}
\end{figure*}

\begin{figure*}
\includegraphics[scale=0.52,trim=1cm 1cm 1cm 3cm]{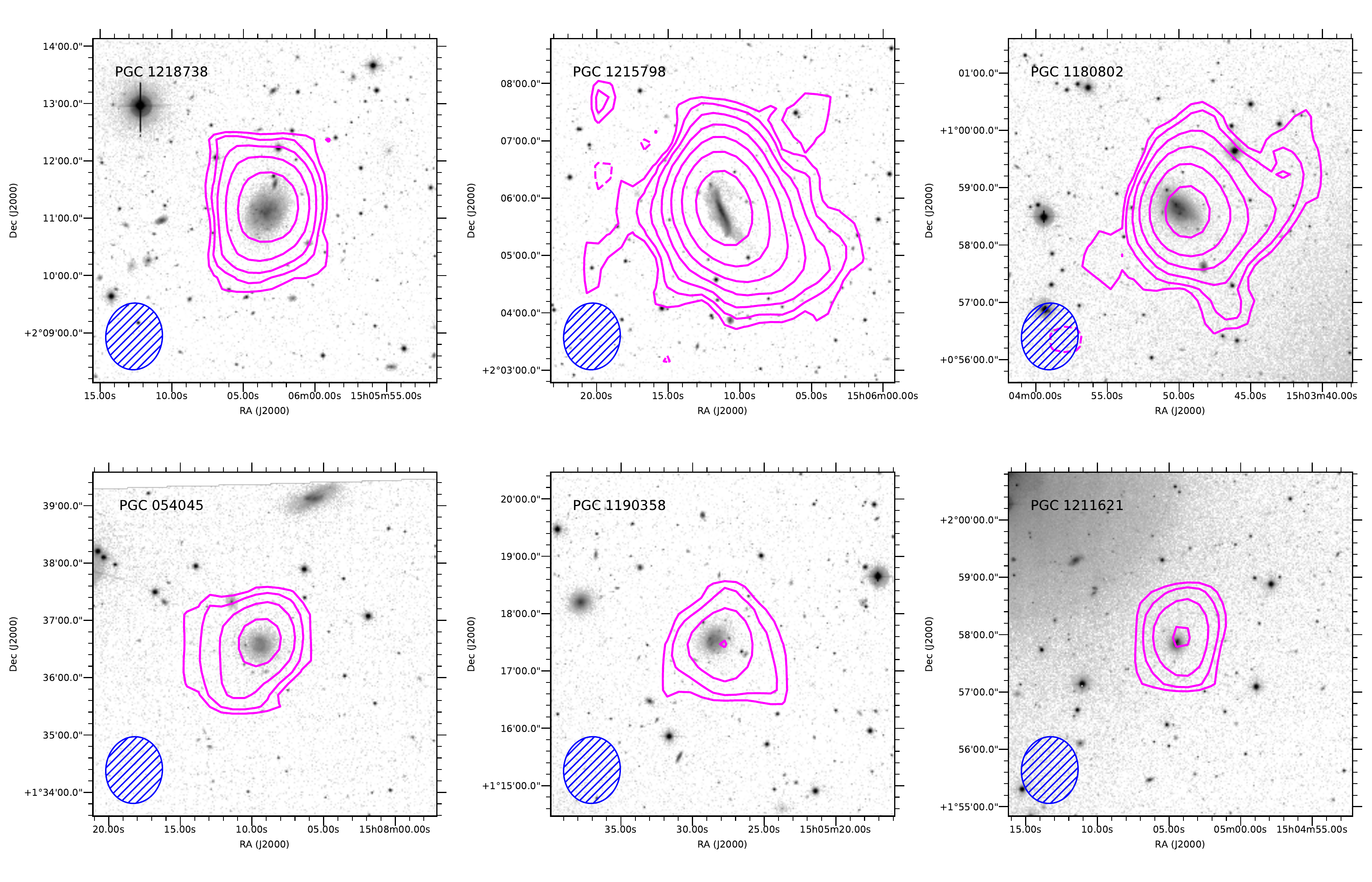}
\caption{\hi-detected galaxies in the NGC~5846 group.  
Contour levels are $(-0.5, 0.5, 1, 2, 4, 8, 16, 32) \times 1.3\times 10^{19}$ \persqcm.
The optical image is the MATLAS $g$ data, except for PGC 054045, where it is SDSS $g$.}
\label{fig:5846m0}
\end{figure*}

\begin{figure*}
\includegraphics[scale=0.9]{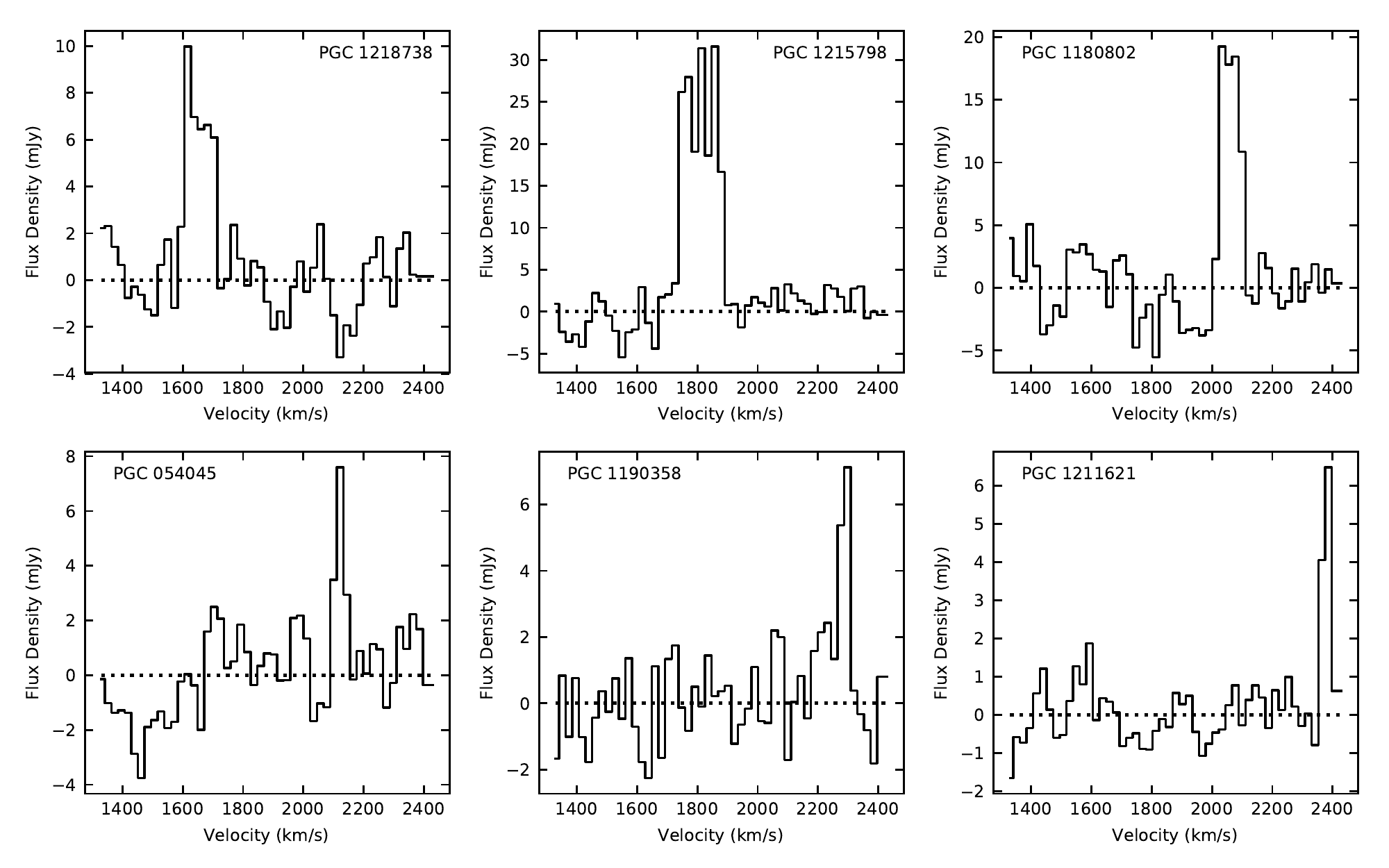}
\caption{Spectra of \hi-detected galaxies in the NGC~5846 group.}
\label{fig:5846spec}
\end{figure*}

Several other galaxies in the vicinity of our targets are also detected in \hi\ emission.
Basic properties are given in Table~\ref{tab:otherhi}; overlays on the optical images are also shown
in Figures~\ref{fig:5577m0}, \ref{fig:others_m0}, and \ref{fig:5846m0}, and spectra are in Figures
\ref{fig:morehispec}, \ref{fig:5577spec}, and \ref{fig:5846spec}.
In the case of the NGC~5846 group mosaic, it is worth noting that
the large spiral NGC~5850 is also in the mosaic field but falls outside of the velocity range covered,
so it is not detected.
Several of the detected galaxies also have previously published \hi\ fluxes, primarily from the Arecibo
and Parkes telescopes, but some appear to be
new \hi\ detections.  The spatial resolution of the VLA data is much better than previous
single-dish data (e.g.\ about a factor of 4 better than Arecibo), so the \hi\ positions are
correspondingly better.  In some cases the resolution also allows us to estimate kinematic position
angles.

The \hi\ properties we measure for NGC~5806, NGC~5811, UGC~09661, and NGC~5577 are all consistent with
measurements published elsewhere, with one
exception.  The \hi\ line flux we measure for NGC~5806 is almost a factor of two lower than previous
single-dish measurements.  The discrepancy is undoubtedly due to the source being 
beyond the half-power point of the VLA primary beam.  The relevant pixels have been multiplied up by
factors of 3 to 5 to account for the primary beam gain, but the imaging quality is still poor that far
out.
For this source, there are better VLA C
configuration data with NGC~5806 at the pointing center \citep{Mundell07}.

PGC~1218738, PGC~1215798, PGC~1180802, PGC~1211621, and AGC~249460 do not have \hi\ detections formally
published elsewhere, but they are listed in the ALFALFA survey data in the first-release version $\alpha$.100 ``A grids" data 
files.\footnote{http://egg.astro.cornell.edu/alfalfa/data/index.php}  Again, the \hi\ properties measured
here are consistent with the single-dish detections, except for PGC~1211621.  In this case the HI
emission is found up to the very last usable channel in the VLA data cube, and it may well extend beyond
that.  The flux measured here is a lower limit for that reason.  

UGC~07945, PGC~054045, and PGC~1190358 do not have any other published \hi\ fluxes, to the best of our
knowledge.  UGC~07945 is outside of the ALFALFA survey area, and the other two are nominally covered by
the ALFALFA survey but their flux is near or below the typical detection limit \citep{alfa40}.

\section{An intriguing isolated \hi\ cloud}

AGC~249460 and the NGC~5574--5576--5577 group (Figure~\ref{fig:5577m0}) merit further discussion.
NGC~5576, NGC~5574, and NGC~5577 form a compact triple with projected
separations $< 200$ \kms\ and $\sim$ 20 and 70 kpc, and the deep image of \citet{duc15} shows dramatic
evidence of an advanced interaction between NGC~5574 and NGC~5576.  However, NGC~5577 appears
undisturbed both in the optical image and in \hi\ emission, so its physical separation from the other two
must be much larger.
Another possible member of the group, AGC~249460, is identified as an \hi\ source in the ALFALFA $\alpha$.100 catalog but an optical counterpart
has not yet been identified.  This makes it an unusual \hi-rich object, as we will discuss in further
detail below.

From the VLA image we estimate the centroid of AGC~249460 to be located at 
14$^h$ 20$^m$ 07.721$^s$, 03\arcdeg 27\arcmin 42.761\arcsec\ (J2000), with a positional uncertainty of about
1.5\arcsec\ in each direction.  This position is 23\arcsec\ east of the one quoted in the ALFALFA
catalog, but that is consistent with the Arecibo positional uncertainties.
We assume a distance of 24.5 Mpc, based on association with NGC~5577, which means the \hi\ flux
corresponds to a mass of 1.3\e{8} \solmass\ or 1.7\e{8} \solmass\ including helium.
Most unusually, the MATLAS image shows no extended optical counterpart at its location, with a limiting
surface brightness in $g$ of 28.5 mag/asec$^2$.  

\subsection{Mass and luminosity}

For an analysis of the physical properties of AGC~249460, it is instructive to compare it to other
optically faint ALFALFA detections that have similar \hi\ fluxes and assumed distances.
Three such objects are discussed by \citet{janowiecki}; one of them
has a faint optical counterpart of about 15\arcsec\ (1.8 kpc) radius, and if we
adopt that same radius here for AGC~249460 we find
that its integrated magnitude must be fainter than $-10.7$ in $g$.
Its luminosity is thus $L_g < 2.2\times 10^6$ L$_\odot$ and its \hi\ mass gives an extremely large
M(\hi)/$L_g > 60$.
The \hi\ mass is similar to the most gas-rich dwarf galaxies of the Local Group, such as NGC~6822, 
but the optical luminosity is similar to that of the faint dwarf spheroidals 
\citep[e.g.][]{mcconnachie}.  Thus, the source has no direct analog in the Local Group, and because of the
fainter surface brightness limit, it has even more extreme M(\hi)/L ratios than the other objects
discussed by \citet{janowiecki}.  This object may be one of the ``darkest" of the dark (optically
faint) galaxies.

The \hi\ emission in AGC~249460 is not well resolved, so dynamical mass estimates have a large
uncertainty, but they do help constrain the nature of the object.  A Gaussian fit to the integrated 
line profile gives a velocity dispersion of 11.0 $\pm$ 1.3 \kms.
A 2D Gaussian fit to the column density gives an estimated {\it deconvolved} size of (43\arcsec\ $\pm$
9\arcsec) by (33\arcsec\ $\pm$ 11\arcsec) FWHM, so a radius of 20\arcsec\ = 2.4 kpc may be reasonable.
These values give a virial mass estimate of 2.7\e{8} \solmass, which is only modestly larger than the
known atomic gas mass of 1.7\e{8} \solmass.
Thus, the data suggest that AGC~249460 is gravitationally bound, but it's not clear whether it requires
a large dark matter component.
AGC~249460 could be a tidal dwarf galaxy with very little dark matter. 
It certainly does not have a dominant stellar component, and it's not entirely clear why there
hasn't been much star formation.  The peak column density in these data is 2.1\e{20}\persqcm, which
suggests that at higher resolution the \hi\ may be thick enough to form H$_2$.

We note also that the estimated virial/gas mass ratio would be larger if the object's actual distance
is substantially less than the 24.5 Mpc we have assumed, and in that case the evidence for dark matter
would be stronger; but at the moment there is no other evidence
against the association with the NGC~5574--5576--5577 group.

\subsection{Origin}

Considering the origin of AGC~249460 and the suggestion that it might be a tidal dwarf, 
we note the obvious interaction between NGC~5574 and NGC~5576 (Figure~\ref{fig:5577m0}).  But the
orientation of NGC~5574's tail suggests that altercation is not relevant for AGC~249460.

Thus, we consider whether AGC~249460 could be gas stripped from NGC~5577, the only other galaxy in the
vicinity that still has HI.
Any presumed interaction must have been weak or 
happened a significant time ago, so that the \hi\ in NGC~5577 should currently appear undisturbed.
For comparison, the orbital time at the outer edge of the \hi\ disc in NGC~5577 is 300 Myr.
The projected distance between the objects is 16\arcmin\
or 110 kpc, and the time required to travel that distance at 300 \kms\ is 360 Myr.
The observed velocity difference between the two is only 10 \kms, though, so that if we presume a
strong interaction then AGC~249460 must be at very nearly the same distance as NGC~5577.
(If it is significantly in front of or behind then the crossing time in the radial direction at 10 \kms\
becomes very long.)

NGC~5577 itself shows possible signs of being stripped; 
\citet{boselli2014} give it an \hi\ deficiency of 0.52 dex compared to its ``normal" analogs.  
Furthermore, its \hi\ mass and $B$ luminosity give M(\hi)/$L_B
\sim 0.16$ \solmass/\solum, which is unusually low for spirals \citep[e.g.][]{janowiecki}.  The maximum
radius of its HI, as measured from the centroids of emission at the extreme velocities, is 50\arcsec\
or only 5.9 kpc.  

All of this evidence is broadly consistent with a picture in which AGC~249460 could
have been a small amount of matter stripped from NGC~5577.  It could also be an unrelated dwarf galaxy
which simply didn't have much star formation.
In either case, the faintness of its optical companion makes it worthy of additional study.

\bsp    
\label{lastpage}
\end{document}